\documentclass[review]{elsarticle}
\usepackage[a4paper]{geometry}
\usepackage{lineno,hyperref}
\usepackage{color,soul}
\usepackage{amsmath}
\modulolinenumbers[5]
\journal{Building and Environment}
\begin{document}
\begin{frontmatter}
\title{Elastic buildings: Calibrated district-scale simulation of occupant-flexible campus operation for hybrid work optimization}

\author[doa,frs]{Mart\'in Mosteiro-Romero\corref{mycorrespondingauthor}}
\cortext[mycorrespondingauthor]{Corresponding author}
\ead{mosteiro@nus.edu.sg}
\author[dob]{Clayton Miller}
\author[dob]{Adrian Chong}
\author[doa]{Rudi Stouffs}

\address[doa]{Department of Architecture, College of Design and Engineering, National University of Singapore, 4 Architecture Drive, SDE1 \#03-01, Singapore 117566, Singapore}
\address[frs]{Singapore-ETH Centre, Future Resilient Systems, CREATE campus, 1 CREATE Way, \#06-01 CREATE Tower, Singapore 138602, Singapore}
\address[dob]{Department of the Built Environment, College of Design and Engineering, National University of Singapore, 4 Architecture Drive, SDE1 \#03-04, Singapore 117566, Singapore}

\begin{abstract}
Before 2020, the way occupants utilized the built environment had been changing slowly towards scenarios in which occupants have more choice and flexibility in where and how they work. The global COVID-19 pandemic accelerated this phenomenon rapidly through lockdowns and hybrid work arrangements. Many occupants and employers are considering keeping some of these flexibility-based strategies due to their benefits and cost impacts. 

This paper explores how demand-driven control strategies in the built environment might support the transition to increased workplace flexibility by simulating various scenarios related to the operational technologies and policies of a real-world campus using a district-scale City Energy Analyst (CEA) model that is calibrated with measured energy demand data and occupancy profiles extracted from WiFi data. These scenarios demonstrate the energy impact of ramping building operations up and down more rapidly and effectively to the flex-based work strategies that may solidify. The scenarios show a 5--15\% decrease in space cooling demand due to occupant absenteeism of 25--75\% if centralized building system operation is in place, but as high as 17--63\% if occupancy-driven building controls are implemented. The paper discusses technologies and strategies that are important in this paradigm shift of operations.

\end{abstract}


\end{frontmatter}


\section{Introduction} \label{sec:introduction}
Demand-driven control strategies in the built environment have been developed and deployed for decades. For example, ventilation rates have long been controlled by demand-driven strategies such as carbon dioxide sensors, from early research projects \cite{Kusuda_1976, Soedergren_1982} to practical implementations since the 1990s \cite{Wang_Jin_1998}. While there is a vast array of potential applications for occupant sensing in buildings, the vast majority of work on occupant sensing and occupant-centric building controls focuses on reducing energy waste and improving occupant comfort with regard to lighting and heating, ventilation and air conditioning (HVAC) \cite{Park_etal_2019, Azimi_OBrien_2022}.

In spite of the huge potential for demand-driven control strategies in buildings, their uptake in HVAC controls has been limited. This is due to the fact that occupant-related aspects in building codes are quite simplistic, and modelers tend to use defaults/code assumptions about occupants to avoid liability, even if they know these values are unrealistic \cite{OBrien_etal_2020_standards}. Hence, ventilation systems are still generally designed to meet buildings' peak design capacity, despite consistent evidence that buildings generally operate far below their design occupancy \cite{Hobson_etal_2020, Duarte_etal_2013, Barbour_etal_2019}. While temperature, airflow, and lighting set points for most commercial facilities can be controlled digitally through the centralized BMS, these set points must still be determined and scheduled manually \cite{Trivedi_etal_2017, Klein_etal_2012}. Hence, there is still a large potential for energy savings from occupancy-based HVAC and lighting controls, ranging from 20\% to 50\% \cite{OBrien_etal_2020_iea}.

%
The problem of building systems being operated by design conditions that do not meet actual building occupancy is exacerbated by the increasing flexibility in working styles, working hours, and work arrangements. Flexible work arrangements are centered on the idea that, rather than bounding people to a fixed desk to carry out their work, they are allowed to choose where they work, either by selecting a location within their usual workplace or by choosing to work remotely from home or other locations, such as coffee shops or libraries. Workplace strategies of this kind are often referred to by a number of terms, such as hot-desking, co-working, desk-sharing, flexible working, office hoteling, and activity-based workspaces \cite{Sood_etal_2020}. Flexible working styles allow employees more freedom to choose their work location, and at the same time desks increasingly no longer belong to a single employee, but may be shared by different occupants \cite{deBakker_etal_2017}. As a result of this shift in how workplaces are used, flexibility has begun to be an issue not just at the zone or even building level, but at the district and urban scale, as building users and of their activities may dynamically change in spatial and temporal distribution.

In early 2020, the COVID-19 pandemic resulted in the restriction of movement of people worldwide, leading to remote working becoming the norm for all non-essential workers. Knowledge and office workers quickly transitioned to working from home full time, mainly individuals from urban areas, with higher paying jobs, benefits, and increased job stability \cite{Peters_etal_2022}. According to an October 2020 survey of 10,332 U.S. adults \cite{Parker_etal_2020}, 71\% of employed participants were working from home compared to 20\% before the pandemic. This exodus from office spaces to the home has shown that such decentralization of office work is possible and desirable in some situations \cite{Sood_etal_2020}.

Given that the push for workspace flexibility precedes the COVID-19 pandemic, remote working arrangements are expected to continue even after the pandemic. There has been a trend towards reducing floor area per person, and many organizations were already implementing shared workspaces \cite{Allvin_etal_2013, Yang_etal_2019} due to the positive cost-benefits to organizations from a reduced office footprint \cite{Marzban_etal_2022}. 25\% of workers in high-income countries are expected to continue remote working either part-time or full-time after the pandemic \cite{Peters_etal_2022}, and the aforementioned survey of U.S. adults found that 54\% of employees would want to work from home all or most of the time after the pandemic \cite{Parker_etal_2020}. Likewise, a survey of 133 U.S. executives and 1200 U.S. office workers found that four in five executives were looking to extend remote work options compared to pre-pandemic periods, while the majority of employees would prefer to be remote for at least three days per week and the majority of executives preferred employees to be in person at least three days per week \cite{PWC_2021}. Enterprises will therefore need to establish what hybrid work environments, models of work and new remote work arrangements will look like in the future \cite{Peters_etal_2022}.

The shift to increased remote working will introduce new challenges to the energy and building sectors. As more and more building occupants adopt flexible working hours, the total scheduled operating time of HVAC systems in commercial buildings could increase \cite{Peng_etal_2017}. 
As workspaces are operated at less than full occupancy and a share of the employees work from home, there is a net increase in the operational floor area per office worker. A study on a planned mixed-use neighborhood in Sweden \cite{Zhang_etal_2020} showed that the electricity demand in the district was largest for scenarios with ``soft'' confinement, where office buildings were assumed to be partially-occupied.

However, supporting working from home and teleconferencing can have strongly beneficial outcomes for emissions if combined with the rationalization and reduction of office space \cite{Kikstra_etal_2021}. There is a need to reconsider how buildings are used and how building systems are operated to avoid energy waste while supporting the needs of occupants who increasingly require flexible working spaces and who may only physically attend their workplace on a part-time basis.

\subsection{Paper scope and structure}

This paper explores how demand-driven control strategies at the district scale might support the transition to increased workplace flexibility through a data-driven simulation approach. In particular, we introduce the notion of \emph{elastic buildings}. Conventional buildings integrate demand responsiveness into zone-based controls. Elastic buildings utilize a larger array of occupant demand-driven strategies at the zone, system, building, and possibly even the district scale. In addition to using occupant sensing technologies, elastic buildings utilize space allocation policies and technologies that pair people with spaces based on their needs. Thus, in addition to demand-driven controls to make buildings reactive to variable occupancy and thus reduce energy demand at the \emph{building} scale, elastic buildings seek to optimize the use of space in order to reduce energy demand at the \emph{district} scale as well.

The advantages of an elastic building system operation are demonstrated by analyzing the effect of different building operation strategies on space cooling demand under different working-from-home (WFH) scenarios. 
Three building operation strategies are considered. The first represents a traditional, centralized building operation, in which buildings are operated according to cooling schedules that affect the entire building. The second strategy represents a fully demand-driven building operation, where buildings are able to provide cooling and ventilation only to occupied spaces, while unoccupied spaces are maintained at the corresponding setback temperature. The third strategy represents an elastic building operation, where spaces are allocated according to demand and buildings are only opened when there is an actual demand for additional workspaces; buildings are assumed to be operated in a way that prioritizes the operation of energy-efficient buildings over buildings with higher cooling demands. 

These scenarios are tested on a case study of a university campus in Singapore, and the space cooling demand for each scenario is estimated through a calibrated campus-scale building energy model that leverages WiFi connection data in order to extract building occupancy patterns at a campus scale.

The paper is organized as follows. Section~\ref{sec:background} gives a general background on demand-driven control strategies in the built environment, flexible work arrangements, and their effects on building energy demand. Section~\ref{sec:methodology} describes the development of the campus-scale building energy demand model, its calibration using building meter and WiFi connection data, as well as the development of occupancy and electricity demand models based on these data. Section~\ref{sec:case_study} describes the case study area and data collection effort. Subsequently, Section~\ref{sec:model} discusses how these data were used to construct the model according to the aforementioned methodology, as well as assessing the model's performance in simulating the baseline case. Section~\ref{sec:results} discusses the simulation results while Section~\ref{sec:discussion} discusses these results' implications on the design of future district energy systems and campus operation.

\section{Background}\label{sec:background}
\subsection{Demand-driven control strategies in the built environment}

Central to the effective implementation of occupancy-driven HVAC controls is information on real-time occupancy and upcoming room occupancy \cite{Peng_etal_2017}. A number of works in the literature discuss the various types of sensors available for occupancy detection \cite{Azizi_etal_2019}, which sensors are best suited for each demand-driven application \cite{Park_etal_2019}, and how data analytics might be implemented in each case \cite{Gunay_etal_2019}. The most often-discussed occupant sensing technologies are motion sensors such as PIR sensors, radio frequency identification (RFID) technology, vision-based sensors (i.e., cameras), ultrasonic sensors, acoustic sensors, environmental sensors (such as CO$_2$ and temperature sensors), and implicit occupancy sensing through energy meters, etc. \cite{Park_etal_2019, Azimi_OBrien_2022, Azizi_etal_2019, Klein_etal_2012, Delzendeh_etal_2017}.

The use of occupancy sensors in commercial lighting control, typically passive infrared (PIR) motion sensors \cite{Naylor_etal_2018}, has become widespread in office and academic settings \cite{Park_etal_2019}. Studies have shown that the implementation of occupancy sensors could help reduce the electricity demand for lighting by between 20--60\%, depending on the configuration, type of space and type of occupancy sensor used \cite{deBakker_etal_2017}, as well as whether the space was irregularly or regularly occupied \cite{Azizi_etal_2019}. Occupancy-on lighting controls, whereby the lights automatically turned on upon occupancy, were once seen as a convenience \cite{OBrien_Gunay_2019}, but can lead to energy waste when sufficient daylight is available \cite{Reinhart_2004}. Occupants are unlikely to turn off the lights when automated controls are present, reducing the energy savings from occupancy sensors by up to 30\% \cite{Sunikka_Galvin_2012}. Including illuminance sensors in addition to occupancy sensors has been found to achieve as much as 65\% energy savings compared to having the lights on all day \cite{Chiogna_etal_2012}. Nagy et al. \cite{Nagy_etal_2015} developed an adaptive control strategy with an illuminance threshold adapted to the preference of the occupants and a time delay setting (i.e., the time before automatically switching off the lights if no detectable occupant movements) adapted to the room's occupancy pattern, and demonstrated up to 37.9\% electricity savings compared to a standard setting control baseline. Due to their effectiveness, numerous building codes give credit to motion sensors that control lighting, while several codes have strict rules against occupancy-on lighting controls \cite{OBrien_etal_2020_standards}.

Similarly, a number of occupant detection technologies and HVAC control strategies have been proposed to reduce energy demand and increase occupants' thermal comfort. In demand-controlled ventilation (DCV), CO$_2$ concentration is used as a proxy for indoor air quality to dynamically control ventilation, and hence save building heating and cooling energy \cite{Lu_etal_2022}. By implementing occupant-driven controls, unnecessary conditioning during vacant hours and over-ventilation due to maximum occupancy assumptions used in ventilation control is avoided \cite{Azimi_OBrien_2022, Esrafilian_Haghighat_2021}. Due to the effectiveness of DCV, numerous building energy codes and standards have recommended or mandated DCV since 1999, though often only under certain circumstances, such as densely occupied spaces \cite{OBrien_etal_2020_standards, Lu_etal_2022}.

For demand response in HVAC control, occupancy detection needs to be accurate, reliable, and able to capture occupancy changes in real time \cite{Erickson_etal_2011}. CO$_2$ sensors can have a time delay due to zone volume, airflow characteristics, and relative distance between sensor and occupant \cite{Wang_Jin_1998}, making them too slow for near real-time HVAC actuation purposes \cite{Balaji_etal_2013}. While PIR sensors, as used for lighting applications, could provide occupant presence information, they are prone to false negatives, thus limiting their applicability for HVAC control, and lighting and HVAC systems are seldom integrated, so their use would incur additional integration costs \cite{Gunay_etal_2019}. Hence, researchers have pursued the use of other types of sensors for occupant detection for building controls.

Agarwal et al. \cite{Agarwal_etal_2010} used a combination of PIR sensors and low-cost magnetic door switches and PIR sensors to detect occupant presence and found potential energy savings of 10--15\% through Energy Plus simulations. Erickson et al. \cite{Erickson_etal_2011} used a vision-based sensor network and found potential annual energy savings of 42\% through simulations. Other researchers have proposed the use of WiFi data as a low-cost source of information on occupant presence, as it relies on pre-existing networks without the need for additional sensors. Occupancy measurements based on WiFi data have shown high accuracy in detection \cite{Melfi_etal_2011}, but occupant counts require an estimate of the relationship between the detected mobile devices and the actual number of people \cite{Park_etal_2022}. Balaji et al. \cite{Balaji_etal_2013} developed a system to infer occupancy from WiFi data and used it to actuate an HVAC system. The system proved to be 86\% accurate in detecting occupancy and had a false negative rate of 6.2\%, leading to 17.8\% lower electricity demands over a single day experiment. The potential to use available networks in lieu of costly sensor installation is especially relevant at the campus scale, and hence Trivedi et al. \cite{Trivedi_etal_2017} proposed a methodology to infer building occupancy from WiFi connection data at campus scale to generate HVAC schedules as inputs to standard building management systems. 

There are, however, some limitations in the use of WiFi data for system controls. These include multiple counting as a result of multiple device ownership, undercounting from occupants without a WiFi enabled device, stationary devices that remain connected for long periods and may be shared by multiple occupants, and positional errors in indoor localization applications \cite{Park_etal_2022}. Furthermore, building systems that operate by a single sensor can easily malfunction if the sensor fails, indicating the low reliability of the system \cite{Azizi_etal_2019}. An increasingly popular approach is the use of sensor fusion techniques and combining different data sources to make use of the advantages of different occupant sensing approaches. Sensor fusion techniques generally achieve higher accuracy than using only a single sensor type \cite{Naylor_etal_2018}, but may require advanced data processing, which may cause a delay in response \cite{Azimi_OBrien_2022}. For example, Hobson et al. \cite{Hobson_etal_2020} present a clustering and motif identification-based approach for day-ahead forecasting of building occupancy patterns using WiFi and electricity demand data. 

The majority of the aforementioned studies can be categorized as reactive controls, in which real-time occupancy detection is used to control building systems. However, such building operation strategies can cause discomfort during transition periods due to the transient nature of thermal and air quality conditions \cite{Hobson_etal_2020, Esrafilian_Haghighat_2021}. In order to overcome this issue, in commercial buildings ventilation is often initiated several hours before expected occupancy to flush out accumulated contaminants and provide thermal comfort by the time of occupant arrival \cite{Azimi_OBrien_2022}. This approach can potentially lead to energy waste, however, as buildings are conditioned while unoccupied. Proactive control approaches \cite{Esrafilian_Haghighat_2021} use learning methods to predict occupant arrival and departure to minimize energy waste while ensuring occupant thermal comfort. For example, Gunay et al. \cite{Gunay_etal_2019} found a potential decrease in heating energy of 30\% and a 13\% decrease in cooling by determining the optimal start and stop times of temperature setback periods. Peng et al. implemented a combination of motion and indoor climate sensors for occupant sensing and an occupancy learning-based demand-driven cooling control that predicts occupants' arrival and duration of stay as well as their temperature setpoint preference based on their past behavior \cite{Peng_etal_2017, Peng_etal_2018}. They tested their approach in a real office environment and found cooling energy savings of 21.2--39.4\% in offices with low to medium occupancy and -6.1--5.7\% in offices with long daily average occupancy durations.

\subsection{Flexible work arrangements and their effects on building energy demand}

The idea of working from home has been proposed for over 50 years \cite{Nilles_1975}. Originally attributed to the oil crisis of the 1970s \cite{Belzunegui_Erro_2020}, the term given to the concept at the time, \textit{telecommuting}, denotes its original motivation: to save on transportation and reduce congestion \cite{Nilles_etal_1976}. By the 1980s, the question of flexible working hours was raised in response to the problem of work-family balance experienced by many workers \cite{Allvin_etal_2013}. The widespread availability of the internet further enabled the shift to increased remote working. Workforce demographic shifts, the increase of knowledge work and the drive of technological evolution have allowed workers to be location-independent \cite{Yang_etal_2019}.

As a result, working from home and other flexible work arrangements have become increasingly common over the past two decades, supported by advances made in information and communications technologies (ICT) \cite{Duarte_etal_2013}. Such arrangements have been found to lead to increased employee satisfaction in an experiment in China \cite{Bloom_etal_2015} and in a U.K. survey \cite{Wheatley_2017}. A review on activity-based work (ABW) found mixed views from workers on the overall experience of working in ABW-supportive environments, but positive to mixed associations between ABW and job satisfaction, work condition and organizational commitment \cite{Marzban_etal_2022}. Benefits for occupants include increasing overall comfort, choice, and control \cite{Engelen_etal_2019} as well as access to an inspiring work environment \cite{Weijs_etal_2019}. Employers, on the other hand, benefit from a healthier and more contented workforce, increased productivity, improved recruitment and retention, reduced absenteeism, and potentially reduced accommodation costs through hot-desking, among others \cite{Wheatley_2017}. 

While working from home has been touted as a means to reduce traffic and energy use, its net beneficial impact on energy demand and greenhouse gas emissions has been questioned due to rebound effects that arise \cite{OBrien_Aliabadi_2020}. Most buildings are not designed to efficiently adapt to variable occupancy \cite{OBrien_Gunay_2019}, and thus tend to waste energy when not fully occupied. A 2014 review found that 26--65\% of building energy use occurred during non-working hours, including nights and weekends \cite{Kim_2014}, an issue that will only be exacerbated if buildings are flexibly occupied most of the time. 

As countries went into full lockdowns in 2020 and nearly all commercial and industrial activities were suddenly halted, energy demand was observed to decrease sharply in most countries where data is available \cite{Krarti_and_Aldubyan_2021, Ruan_etal_2020}. The effects of the pandemic on energy demand varied by sector; however, while commercial activity decreased, people spent an increasing amount of time at home. Fully enforced lockdowns and stay-home orders in 2020 increased the residential sector energy demand by a range from 11\% to 32\% for several countries with available metered data, mainly driven by HVAC and appliance use during daytime hours \cite{Krarti_and_Aldubyan_2021}. Similar effects were observed in Canada \cite{Rouleau_and_Gosselin_2021}, the United States \cite{Kawka_and_Cetin_2021}, and South Korea \cite{Kang_etal_2021}. In Singapore, the electricity demand in households went up as a result of government-mandated lockdowns, after which a ``new normal'', in which people stayed in most of the time, was achieved \cite{Raman_and_Peng_2021}.

The rise in residential energy demand due to the increase in remote working was not necessarily accompanied by a proportional decrease in demand in non-residential buildings. In an Australian university campus, the HVAC systems for academic, administration, retail and teaching buildings were shut down for 9--10 weeks during the pandemic, leading to 44--57\% savings in total energy demand \cite{Gui_etal_2021}. However, a university in the Netherlands reports a decrease of only 3\% during COVID-19 even though occupancy was only 25\% \cite{tudelft_web}. In such campus buildings, a weak correlation between occupancy rate and energy usage has been found, possibly due to the use of fixed HVAC operation schedules \cite{Anand_etal_2022}. This could help to explain why the energy demands during the pandemic were not proportionally lower in spite of the decrease in or complete absence of building occupants. Indeed, while the global final energy demand decreased, the buildings sector showed a 3\% increase in final energy demand globally in 2020 \cite{Kikstra_etal_2021}.

In the context of increased workspace flexibility, therefore, there is a need to plan buildings and districts that are more reactive to occupants' needs in order to avoid the types of energy waste observed before and during the COVID-19 pandemic.

\section{Methodology}\label{sec:methodology}
\begin{figure}
  \includegraphics[width=\linewidth]{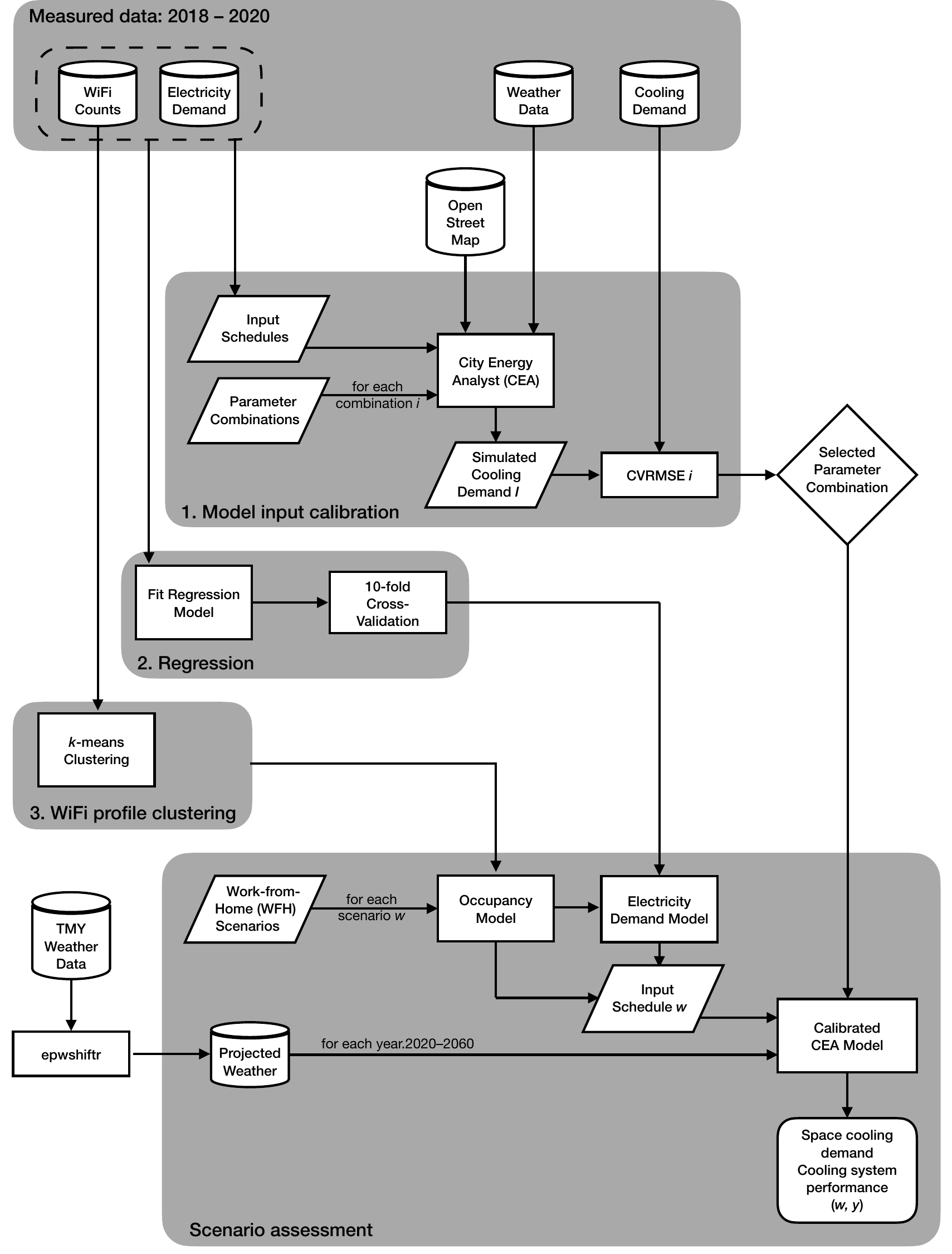}
  \caption{Workflow used to develop the district energy demand model and simulate the application scenarios.}
  \label{fig:workflow}
\end{figure}

This paper analyzes the performance of different building operation strategies under a range of work-from-home (WFH) scenarios for present-day and future climates through district-scale energy demand simulations. The workflow used to develop the energy demand model for the case study area and to apply it for scenario assessment is shown in Figure~\ref{fig:workflow}. 

Building energy demand modeling was carried out on City Energy Analyst (CEA), an open-source tool for urban building energy modeling and supply system optimization \cite{cea_team_2021}. Each building's footprint was extracted from OpenStreetMap \cite{OpenStreetMap} and extruded by the building's number of floors in order to obtain the building volume. CEA includes archetype databases for Switzerland and Singapore that assign typical building properties based on the typology of a given building \cite{Fonseca_Schlueter_2015}. These include both building materials and their thermal properties as well as power densities for lighting and appliances and energy system operation set points.

In order to improve the model accuracy, these properties were then adjusted through model calibration. For this purpose, hourly measured energy demand, weather and WiFi connection data were collected and used as inputs in the model. Due to the case study's location in Singapore, the energy meter data collected is limited to building cooling and electricity demand data, as no heating is necessary. In order to analyze the effects of changes in building occupancy on the electricity demands of the district, a regression model was fit based on the aforementioned hourly data. A baseline occupancy scenario was finally developed by clustering WiFi connection profiles for each building as an input to the model. This calibrated building energy demand model was then used to assess different work-from-home (WFH) and future weather scenarios.

\subsection{Energy demand model calibration}
The variables to be calibrated were selected based on a previous sensitivity analysis of the CEA thermal loads model \cite{MosteiroRomero_etal_2017}, in which the set point and setback temperatures of the cooling systems, the window-to-wall ratios, and the windows' thermal transmittance (U-value) and solar transmittance (g-value) were found to be the driving factors for the deviation in the cooling demand results of the model. While some of these properties could, in principle, be obtained through building plans, material specifications and measured data about system operation parameters, at the district-to-urban scale such information is typically not available for individual buildings, and hence calibration can help improve model accuracy. For simplicity, the model was calibrated by brute force optimization using discrete changes in the variables chosen. The selected variables and their distributions are shown in Table~\ref{tab:calibration}. In addition to these parameters, different types of window glazing properties (Table~\ref{tab:calibration_windows}) were also considered. In order to avoid selecting unrealistic or infeasible material properties, these were selected from the CEA materials database for the Singaporean context \cite{cea_sg_database}.

\begin{table}[htbp]
  \centering
  \small
    \caption{Variables chosen for the calibration and their distributions.}
    \label{tab:calibration}
    \begin{tabular}{ l c c c }
    \hline
     Parameter                                          &   Minimum &   Maximum &   Increment   \\
    \hline
     Cooling set point temperature [$^\circ$C]          &   17      &   29      &   1           \\
     Cooling setback temperature ${}^\star$ [$^\circ$C] &   17      &   29      &   1           \\
     Window-to-wall ratio [-]                           &   0.29    &   0.89    &   0.1         \\
     Infiltration rate [ach]                            &   0.1077  &   0.6463    &   0.1077    \\
    \hline
    \multicolumn{4}{l}{${}^\star$ No cooling system operation during unoccupied hours was also considered.}\\
    \end{tabular}
\end{table}

\begin{table}[htbp]
  \centering
  \small
    \caption{Windows taken from the CEA materials database.}
    \label{tab:calibration_windows}
    \begin{tabular}{l c c c }
    \hline
     Description        & Thermal transmittance & Solar transmittance & Emissivity \\
     ~              
                        & (U-value)             & (g-value)           & ~          \\
    \hline
     Single glazing     &   5.4    &   0.85     &   0.89    \\
     Double glazing 1   &   1.8    &   0.29     &   0.89    \\
     Double glazing 2   &   2.2    &   0.2      &   0.89    \\
    \hline
    \end{tabular}
\end{table}

For each of the resulting input parameter combinations, a yearly energy demand simulation was carried out using the CEA demand model for the case study area. During the calibration procedure, measured electricity demand data was used as an input to the thermal demand model. Hourly WiFi connection counts for the same period as the measured electricity demand data were used as occupancy schedules for the thermal model. Finally, measured weather data from the same period was used in the simulations. 

Each combination of input parameters was then assessed according to their normalized mean bias error (NMBE) and coefficient of variation of root-mean-square error (CV(RMSE)). Since the goal of this step was to calibrate the thermal properties of the building, the hourly cooling demand was used for assessing the results for each combination. For each building, the combination of parameters resulting in the lowest CV(RMSE) for the hourly cooling demand with respect to the measured data was selected. The calibrated model's performance was then assessed based on the threshold values specified in ASHRAE Guideline 14 \cite{ASHRAEGuideline14}, that is 
NMBE below 10\% 
and CV(RMSE) below 30\% for hourly calibration data.

\subsection{Regression model of building electricity demand}
The scenarios analyzed in this paper relate to partial building occupancy due to flexible work arrangements. Hence, in order to account for the effects of occupant presence on the electricity demands of the buildings in the area, a regression model was fitted for each building. This model is based on the assumption that the buildings' electricity demand can be split into base loads, occupant-dependent loads, occupant-independent lighting loads, and ventilation loads as follows:
\begin{equation}
    \Phi_{el} = \Phi_{el, bl} + \Phi_{el, occ} + \Phi_{el, l} + \Phi_{el, v} \label{eq:electricity_tot}
\end{equation}

The base loads $\Phi_{el, bl}$ (W) are assumed to be constant. The occupant-driven loads $\Phi_{el, occ}$ are assumed to be directly proportional to the number of occupants $N_{occ}$:
\begin{equation}
    \Phi_{el, occ}(t) = \phi_{occ} \cdot N_{occ}(t) \label{eq:electricity_occ}
\end{equation}
where $\phi_{occ}$ is the demand per occupant (W/p).

The lighting loads $\Phi_{el, l}$ (W) are assumed to follow a double logistic function, rising in the morning at an initial time $t_{0}$ and decreasing in the evening at a final time $t_{1}$:
\begin{equation}
    \Phi_{el, l}(t) = \phi_{l} \cdot \left[s \cdot w(t)\right] \cdot \left[\left(1 + e^{-k \cdot (t - t_{0})}\right)^{-1} - \left(1 + e^{-k \cdot (t - t_{1})}\right)^{-1}\right] \label{eq:electricity_l}
\end{equation}
where $\phi_{l}$ (W) is the peak occupant-independent load for lighting, $s$ is a constant to account for the difference in magnitude on weekdays and weekends, $w(t)$ is equal to 1 on weekdays and 0 on weekends, and $k$ is the logistic growth rate.

Finally, during extended unoccupied times, concurrent electricity and cooling loads were observed as buildings were flushed once a day. Hence an electricity demand due to ventilation was also fit as a function of the cooling demand:
\begin{equation}
    \Phi_{el, v}(t) = \phi_{v} \cdot \Phi_{cool}(t) \label{eq:electricity_v}
\end{equation}
where $\Phi_{cool}$ (W) is the cooling demand at time $t$, and $\phi_{v}$ is the electricity load due to building ventilation as a function of the cooling load.

The regression parameters ($\Phi_{el, bl}$, $\phi_{occ}$, $\phi_{l}$, $\phi_{v}$, $s$, $k$, $t_{0}$ and $t_{1}$) were fit by using the available measured data for the electricity and cooling demand and the hourly WiFi connections of each building, with 10-fold cross-validation. These resulting parameters were then used to define the input schedules to the energy demand model. 

\subsection{Occupant modeling}
Realistic occupancy schedules for the buildings in the case study area during normal operation were developed as a baseline. For this, building occupancy patterns were extracted from available data on the WiFi connections in each building in the case study area during the same period as the measured electricity and cooling demand data. The data available took the form of unique device counts at each hour of the year for each building in the case study and was previously shown to improve the model's prediction accuracy \cite{Chong_etal_2021_AE}.

While the number of devices connected to the WiFi network is an indicator of the number of occupants in a building, occupants may have zero or multiple WiFi-connected devices at their disposal. In this step, occupancy \textit{patterns} were extracted from the WiFi data, which took the form of a value between 0 and 1 at each time of the day. These were then multiplied by the maximum number of occupants in each building, calculated based on the occupant density and floor area for the corresponding building use type.

In order to obtain typical occupancy patterns from the WiFi connection data, an approach similar to Zhan and Chong's \cite{Zhan_and_Chong_2021} was followed. The data was first normalized by subtracting the minimum and maximum value in order to display all profiles within a [0, 1] range:
\begin{equation}
    n_{devices}'(t) = \frac{n_{devices}(t) - \text{min}\left(n_{devices}\right)}{\text{max}\left(n_{devices}\right) - \text{min}\left(n_{devices}\right)}
\end{equation}
$k$-means clustering was then applied in order to identify typical occupancy patterns for each building. Each day in the dataset was assigned an occupancy pattern cluster, and subsequently a ``weekday'', ``Saturday'' and ``Sunday'' schedule was assigned to each week in the year for each building in the area. These ``typical'' schedules for each building were then assigned as the baseline occupancy schedules for the building energy model. The schedules for different occupancy scenarios were then adjusted only by changing the peak number of occupants in each scenario:
\begin{equation}
    n_{occupants}(t) = n_{devices,i}'(h) \cdot \left(\sum_{f} \frac{A_{f}}{\delta_{f}}\right) \cdot \left(1 - WFH\right)
\end{equation}
where $n_{occupants}(t)$ is the number of occupants at a given time step $t$, $n_{devices}(h, i)$ is the number of devices at the given hour of the day $h$ for the assigned cluster $i$, $A_f$ and $\delta_f$ are the floor area share and occupant density for building use type $f$ in the current building, and $WFH$ is the share of people working from home in any given scenario.

\subsection{Scenario assessment} \label{sec:scenarios}

\begin{figure}
  \includegraphics[width=\linewidth]{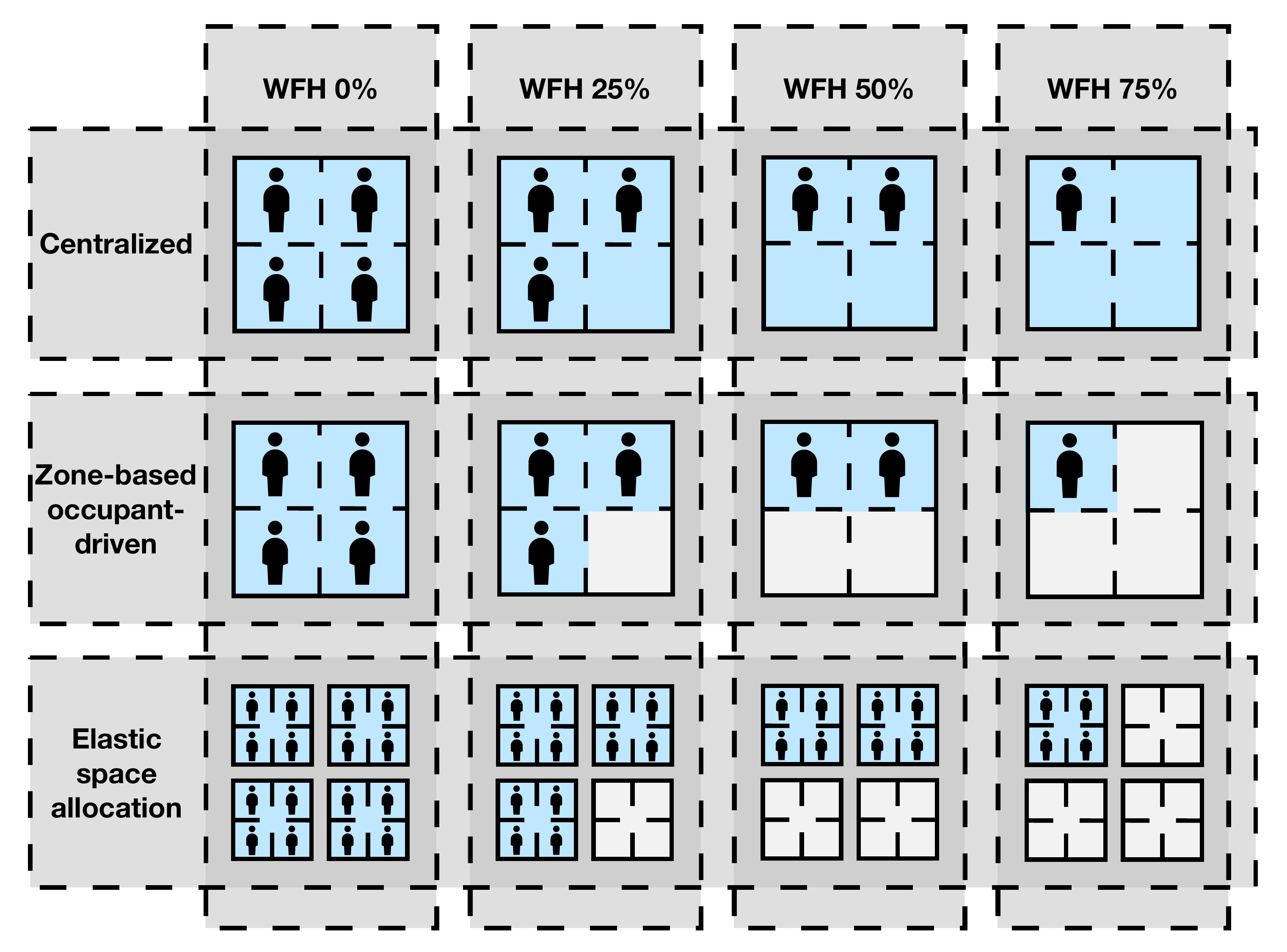}
  \caption{Scenario matrix showing the different work-from-home (WFH) shares assumed and the different building operation strategies. Each solid box represents one building, and the dashed lines represent subzones within the building.}
  \label{fig:scenario_schematic}
\end{figure}

The calibrated building energy demand model with realistic occupancy patterns and associated electricity demands was then used for scenario evaluation. Different scenarios were defined in order to explore the long-term effects of the increase in remote working and studying as a result of the COVID-19 pandemic. Figure~\ref{fig:scenario_schematic} shows the scenarios defined to assess the effects of different working-from-home scenarios on space cooling demand under different building operation strategies. Four different ``work-from-home'' cases were considered (0\%, 25\%, 50\% and 75\%) and in each case, the predefined percentage of students and employees were assumed to stay at home, consequently reducing the occupancy in university and office buildings. 

Furthermore, three different building operation strategies were assumed for each of these cases in order to assess how building operation might affect the cooling demands of the case study area in each scenario. In the \emph{centralized} case, each building's cooling systems can only be turned on or off centrally, and therefore the entire building is conditioned whether it is fully occupied or not. In the \emph{zone-based, occupant-driven} case, only occupied spaces are conditioned, such that the conditioned floor area is equal to the share of the occupants working or studying on-site. While reducing space cooling demand, in this case, a significant share of buildings' floor areas is left unoccupied most of the time. In order to support the rationalization and reduction of office space, the final building operation strategy aims for \emph{elastic space allocation} in buildings, whereby building occupants would need to book a workspace in advance (for example, through a mobile application). Thus, partial building occupancy would be avoided by prioritizing full occupancy in the most energy-efficient buildings, while 
all other buildings are assumed to be closed.

Part load operation, as introduced in the zone-based occupant-driven scenario, can affect system operation, reducing the coefficient of performance (COP) of the cooling systems and thus leading to higher electricity loads for building operation. In order to investigate this effect, the performance of the cooling systems in the area under each building operation scenario was also considered. While buildings in the area are cooled by different centralized chiller plants, for comparison's sake, here we assume each building is equipped with its own cooling systems. This would be a typical arrangement for a district where buildings were not owned by the same stakeholder. Building-scale chillers were assumed to be sized according to the baseline cooling demands, with an assumed safety factor of 15\% \cite{COMNET}:
\begin{equation}
    Q^{rated}_{chiller,k} = P^{0}_{cooling, k} \cdot 1.15
\end{equation}
where $Q^{rated}_{chiller,k}$ is the rated capacity of the chiller installed in building $k$ and $P^{0}_{cooling,k}$ is the peak cooling power of building $k$ in the baseline year.

The coefficient of performance (COP) for part-load operation was calculated as follows:
\begin{equation}
        COP = 0.4 \cdot \frac{T_{supply}}{T_{return} - T_{supply}} \cdot PLF
\end{equation}
where $T_{supply}$ and $T_{return}$ are the supply and return temperature of the cooling system (assumed to be 13$^\circ$C and 6.5$^\circ$C, respectively, according to the CEA database) and $PLF$ is the part load factor, calculated according to an empirical equation \cite{COMNET}:
\begin{equation}
    PLF = 0.33018833 + 0.23554291 \cdot PLR + 0.46070828 \cdot PLR^2
\end{equation}
where $PLR$ is the part load ratio:
\begin{equation}
    PLR(t) = \frac{P_{cooling}(t)}{Q^{rated}_{chiller}}
\end{equation}

In addition to the effects of flexible work arrangements, the long-term effects of climate change on the case study area's projected cooling demands and building system operation were also considered. For this, future weather files were created using the R package \textit{epwshiftr} \cite{epwshiftr}. The package takes a weather file as an input and generates weather projections for future years based on the Coupled Model Intercomparison Project Phase 6 (CMIP6) data \cite{Eyring_etal_2016}. For this case study, CMIP6 model AWI-CM-1-1-MR was selected together with scenario SSP585, which represents the high end of the range of future pathways in the integrated assessment model literature and is planned to be used by a number of CMIP6-endorsed projects to help address their scientific questions \cite{ONeill_etal_2016}.

\section{Case study} \label{sec:case_study}
\begin{figure}
  \includegraphics[height=0.49\linewidth]{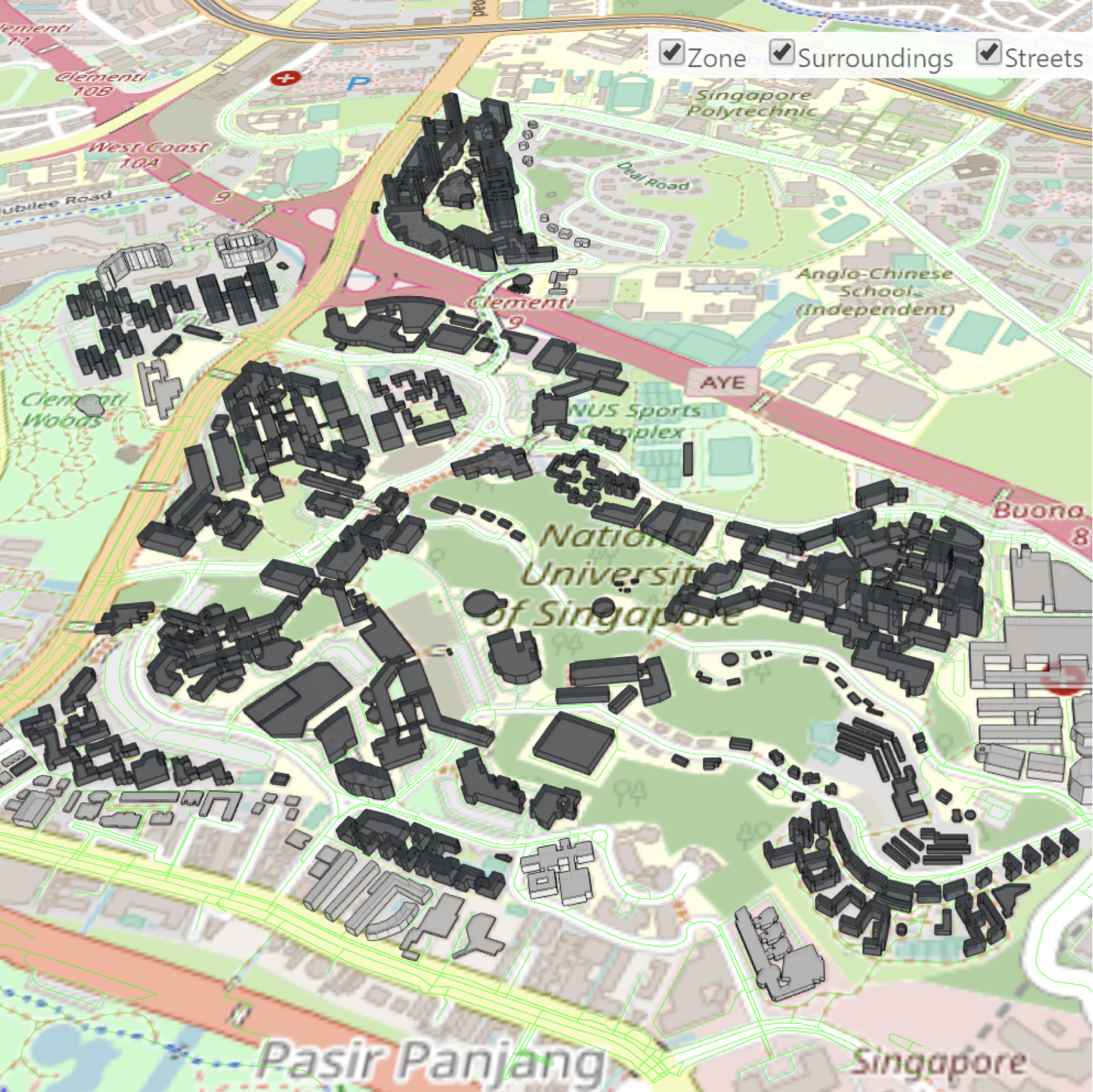}
  \includegraphics[height=0.49\linewidth]{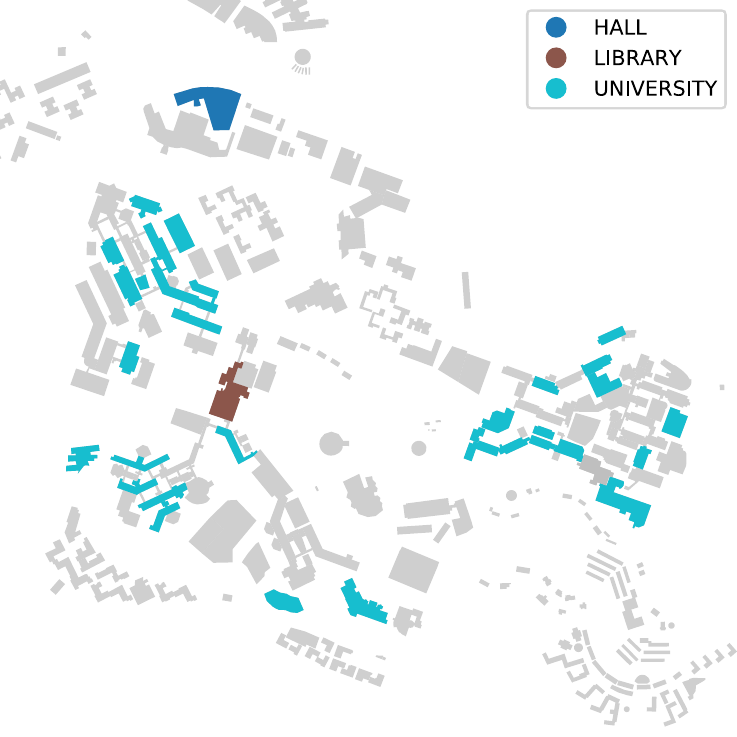}
  \caption{View of the university campus model in CEA (left) and selected buildings in the present case study (right).}
  \label{fig:case_study}
\end{figure}

The scenarios are assessed through the case study of a university campus in Singapore (Fig.~\ref{fig:case_study}). The campus comprises mixed-use buildings including classrooms, offices, laboratories, residential areas, and other amenities. Due to Singapore's tropical climate, the buildings' energy demands correspond mainly to cooling (which the buildings get from centralized chiller plants) and electricity, with seasonal changes in the demands mainly driven by occupancy (\textit{i.e.}, by the university calendar) rather than changes in weather patterns.

The data collected on-site comprises hourly electricity and cooling meter data, as well as WiFi connection logs for each building for the period 2018--2020. In addition to this data, measured weather data for the same period was also collected from the weather station at Singapore's Changi Airport. 
The available WiFi data took the form of hourly values for the number of devices connected to a given building's WiFi access points during the same three-year period. Due to an issue with the WiFi logging system, however, no counts are available for the first half of 2020. 

The building energy model presented in this paper focuses on 35 buildings, as shown in Figure~\ref{fig:case_study}, which are the only buildings for which electricity and cooling demand, as well as WiFi data, was available. These correspond to 33 mixed-use university buildings, one library, and one performance hall. Within this dataset, energy demand and WiFi data for the full three-year time period were only available for 12 buildings, while for 22, the data was restricted to 2018 and 2019, and for 1 only 2020 data was available. 

\begin{figure}
  \includegraphics[width=\linewidth]{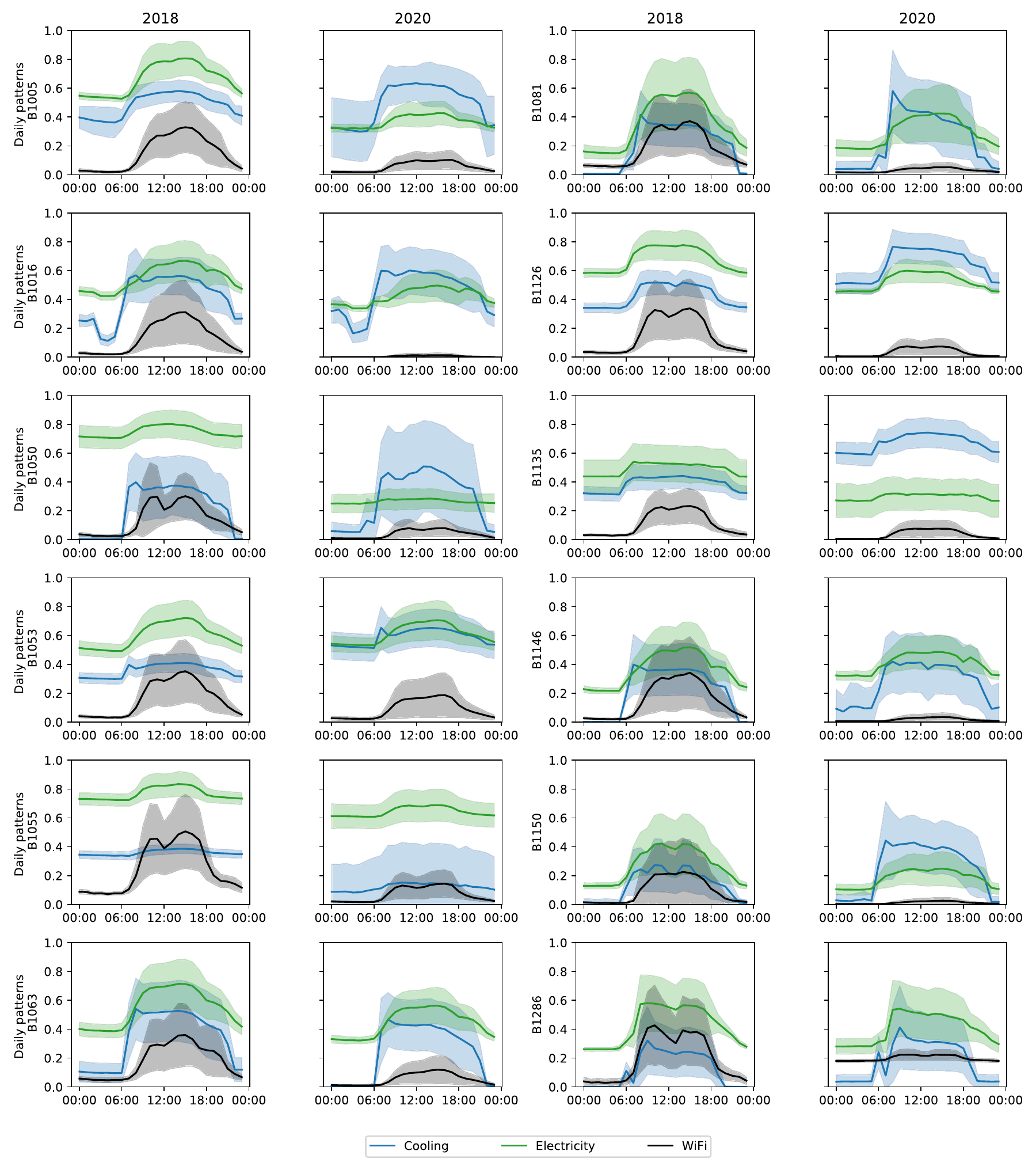}
  \caption{Mean daily cooling and electricity demand and WiFi connection profiles for 2018 and 2020, the shaded areas correspond to one standard deviation in the dataset.}
  \label{fig:data_2018_2020}
\end{figure}

The daily cooling demand, electricity demand, and WiFi connection profiles for 2018 and 2020 are shown in Figure~\ref{fig:data_2018_2020}. The demand profiles are normalized by the peak value from 2018 to 2020 in order to display all profiles within a [0, 1] range. The WiFi device profiles are noticeably lower for all buildings except B1286, where WiFi-connected sensors were installed sometime between 2019 and the beginning of 2020. For most buildings, the cooling demand profile remains fairly constant, with the exception of buildings B1053, B1081, B1126, B1146 and B1150, for which the cooling demand was noticeably higher in 2020. The cooling demand in 2020 was only significantly decreased for building B1055. The daily electricity demand patterns for most buildings remained relatively constant, although there was a noticeable decrease for buildings B1005, B1055, B1081, B1126 and B1135.

\begin{figure}
  \includegraphics[width=\linewidth]{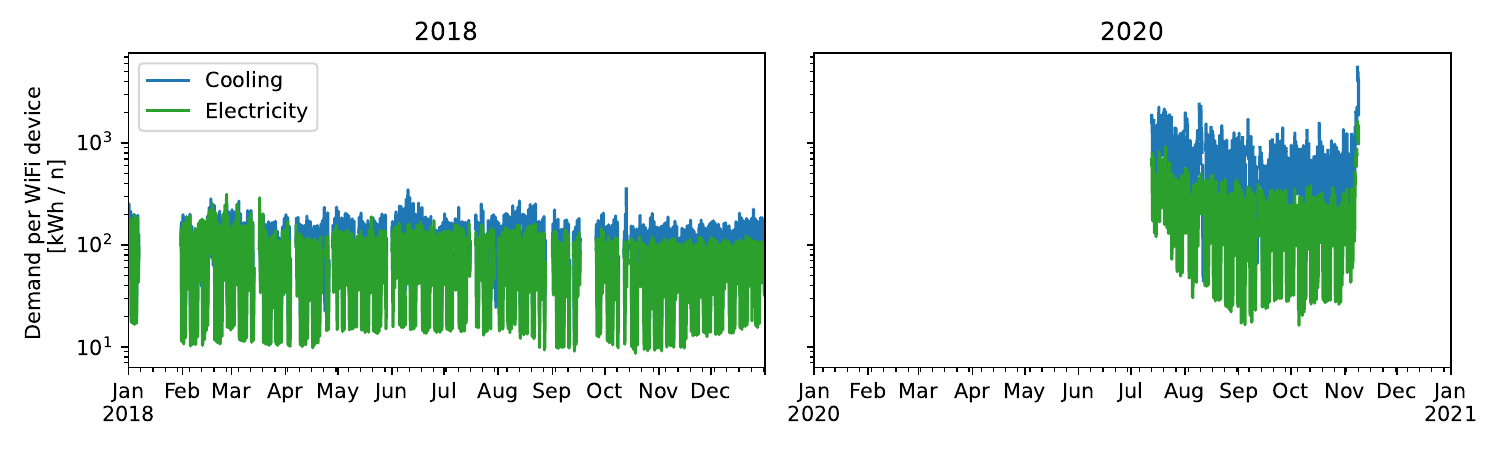}
  \caption{Total cooling and electricity demand normalized by the number of WiFi-connected devices at each hour for 2018 and 2020 for all buildings in the case study.}
  \label{fig:data_2018_2020_per_device}
\end{figure}

As a result of the demand and occupancy patterns observed above, the electricity and cooling demand per WiFi-connected device (an indicator of the change in demand per building occupant) increases significantly, as shown in Figure~\ref{fig:data_2018_2020_per_device}. This would appear to indicate that building systems on campus were operated normally during the COVID-19 pandemic. The increased ventilation requirements, as well as the increase in the space required per occupant due to social distancing, led to increased cooling demand. If remote learning and flexible work arrangements do indeed become a long-term feature in university campuses, however, building operations should be adapted to better provide for occupants' needs while minimizing energy use.

\section{Model construction and assessment} \label{sec:model}
\subsection{Energy demand data preprocessing}
\begin{figure}
  \includegraphics[width=\linewidth]{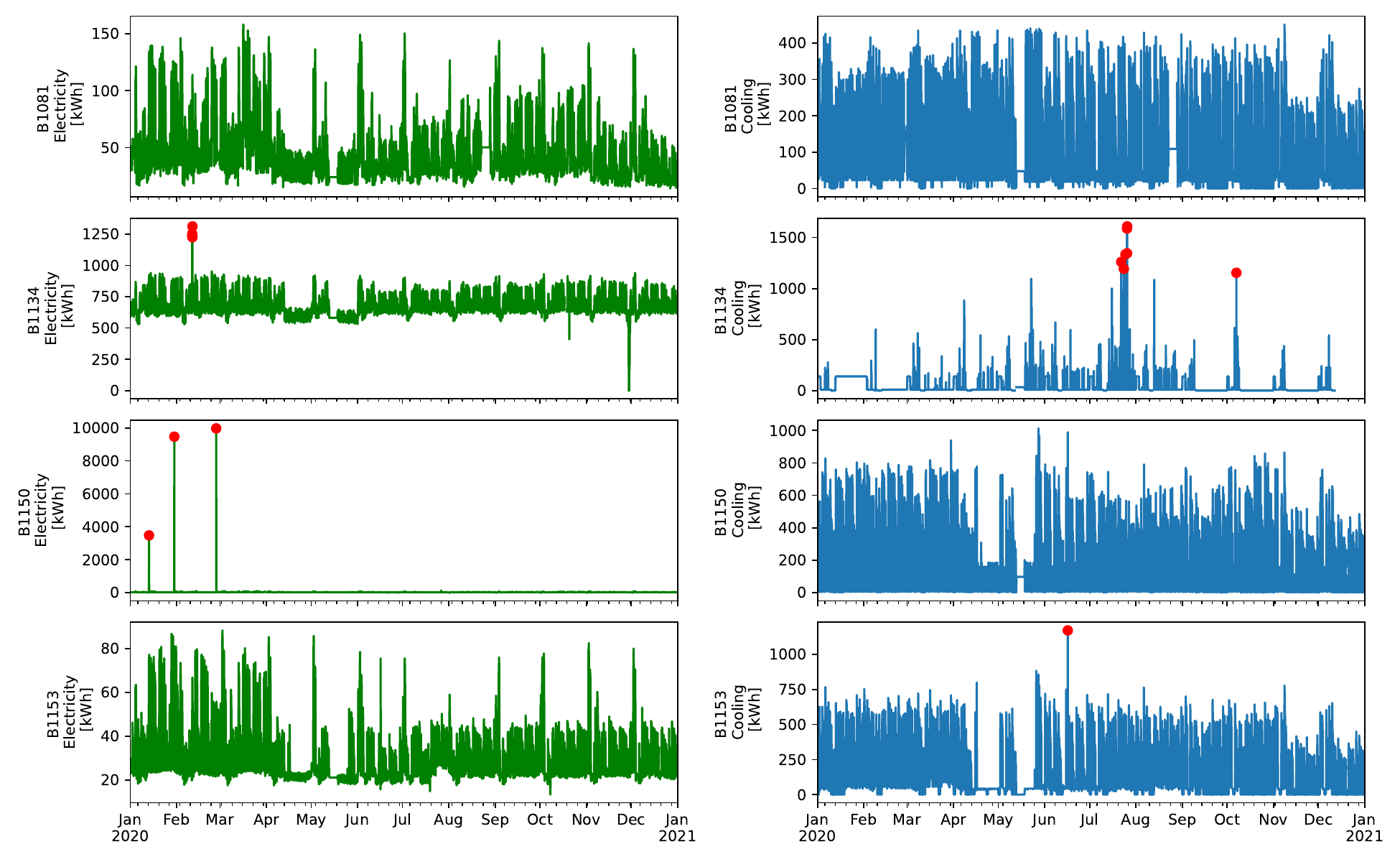}
  \caption{Electricity and cooling demands for selected sample buildings showing outliers in the measured data (in red).}
  \label{fig:outliers}
\end{figure}

For the model development, measured hourly cooling and electricity demand data were available for a subset of the buildings in the case study area collected between 2018 and 2020. The electricity and cooling data needed to be preprocessed in order to clean up metering errors such as the ones seen in Figure~\ref{fig:outliers}. These outliers were detected by their Z-score:
\begin{equation}
    Z = \frac{X - \mu}{\sigma}
\end{equation}
Data points with a Z-score greater than five were recursively removed until no value in the dataset for a given building exceeded this threshold. 

The meter data was used in the building energy model development as discussed in Section~\ref{sec:methodology}. Since the buildings in the case study area include a variety of research spaces, the base loads for cooling and electricity account for a significant share of the demands (e.g., B1134 in Figure~\ref{fig:outliers}). Since the calibration procedure aims to estimate the space cooling demand of the buildings and not the specific research equipment housed within them, the base loads were not considered as part of the calibration procedure. A given building's space cooling systems were therefore assumed to be on only when the demand for cooling exceeded this base load.

\subsection{Cooling model calibration}
\begin{figure}
  \includegraphics[width=\linewidth]{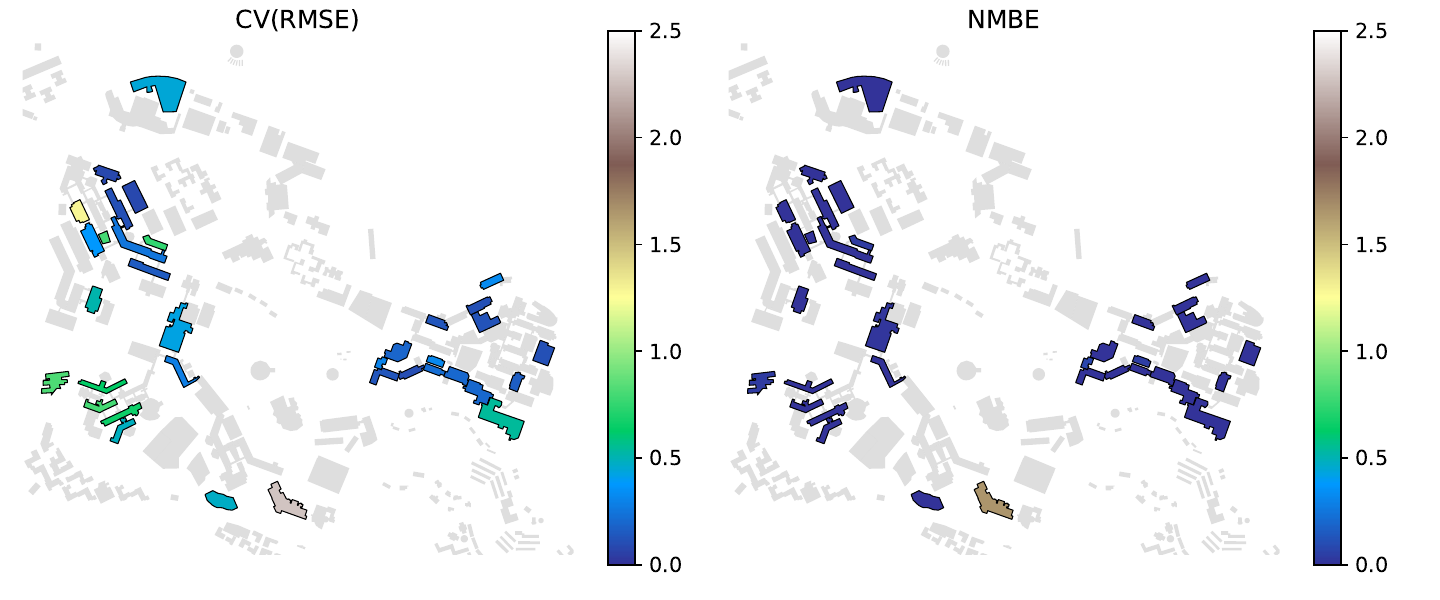}
  \caption{CV(RMSE) and NMBE for the hourly cooling demands predicted for each building in the case study after calibration.}
  \label{fig:calibration_map}
\end{figure}

\begin{figure}
  \includegraphics[width=\linewidth]{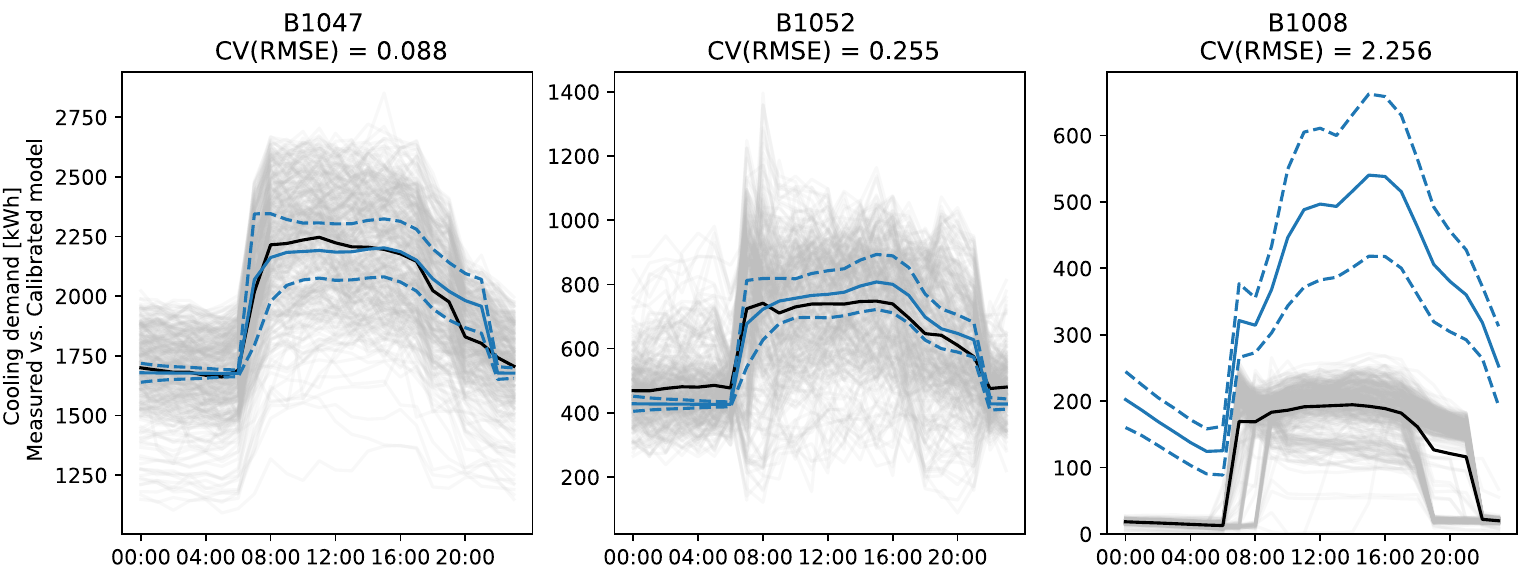}
  \flushright
  \includegraphics[width=0.25\linewidth]{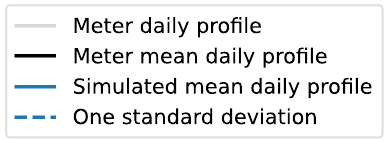}
  \caption{Calibrated cooling model compared to the measured cooling demand data for three sample buildings: two that are successfully calibrated and one that is not.}\label{fig:measured_vs_calibration}
\end{figure}

After calibrating the buildings' thermal properties and building system operation parameters, about half of the buildings' cooling demand models (18 out of 35) lie within the accepted threshold for hourly building energy model calibration of 30\% CV(RMSE) while the results for 34 out of 35 buildings lie within the 10\% NMBE threshold (Fig.~\ref{fig:calibration_map}). The daily cooling demand patterns for three sample buildings are shown in Figure~\ref{fig:measured_vs_calibration}. Both sample buildings whose CV(RMSE) falls within the calibration threshold show a fairly good agreement in the average daily patterns. 

The example building with the highest CV(RMSE) shows the measured cooling loads are always much lower than those predicted by the calibrated model. This appears to indicate a building that is beyond the calibration capability of the approach used. Common causes for such a discrepancy are a mixed-use building being mislabeled as monofunctional with a highly energy-intensive building use type, or the metered data corresponding to only a portion of the building \cite{Sun_etal_2022}.

In order to avoid misrepresenting the potential savings observed in the different scenarios, the results discussed in Section~\ref{sec:results} refer exclusively to those buildings for which the model's CV(RMSE) was within the acceptable threshold.

\subsection{Electricity demand regression model}
\begin{figure}
  \includegraphics[width=\linewidth]{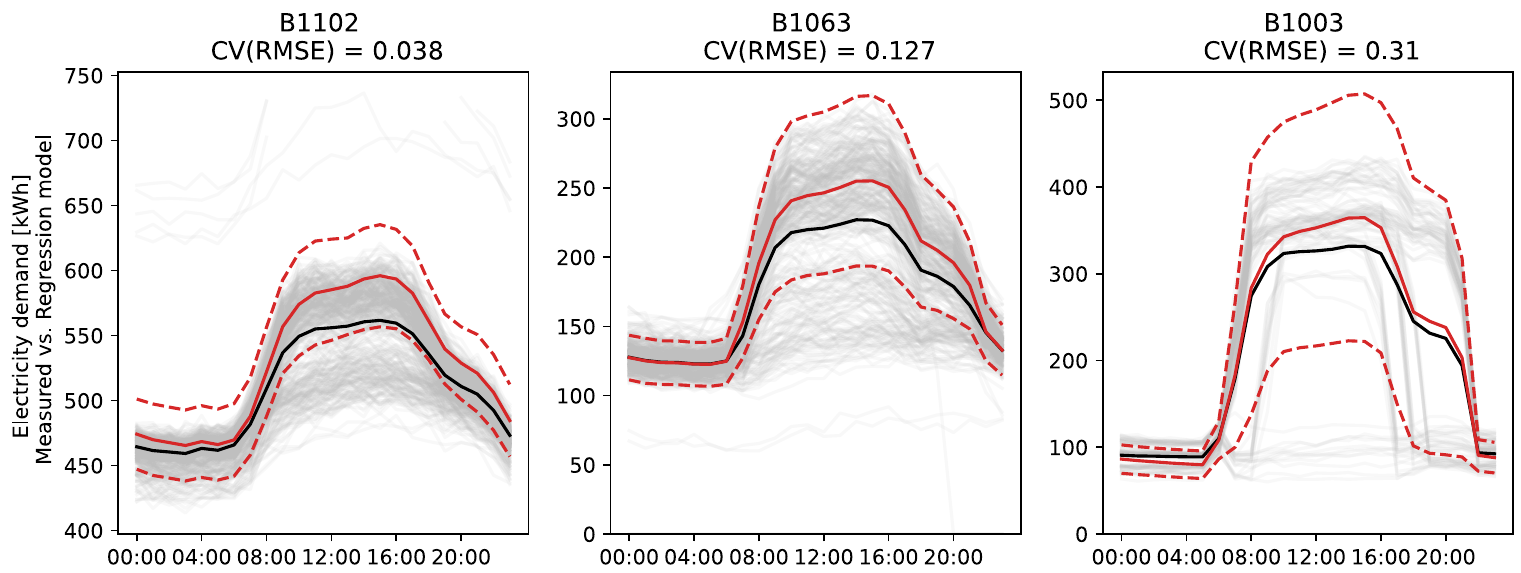}
  \flushright
  \includegraphics[width=0.25\linewidth]{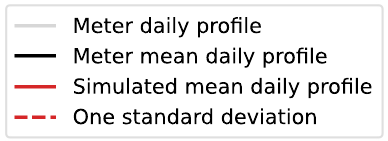}
  \caption{Electricity regression model compared to the measured electricity demand data for three sample buildings.}
  \label{fig:electricity_regression}
\end{figure}

\begin{figure}
  \includegraphics[width=\linewidth]{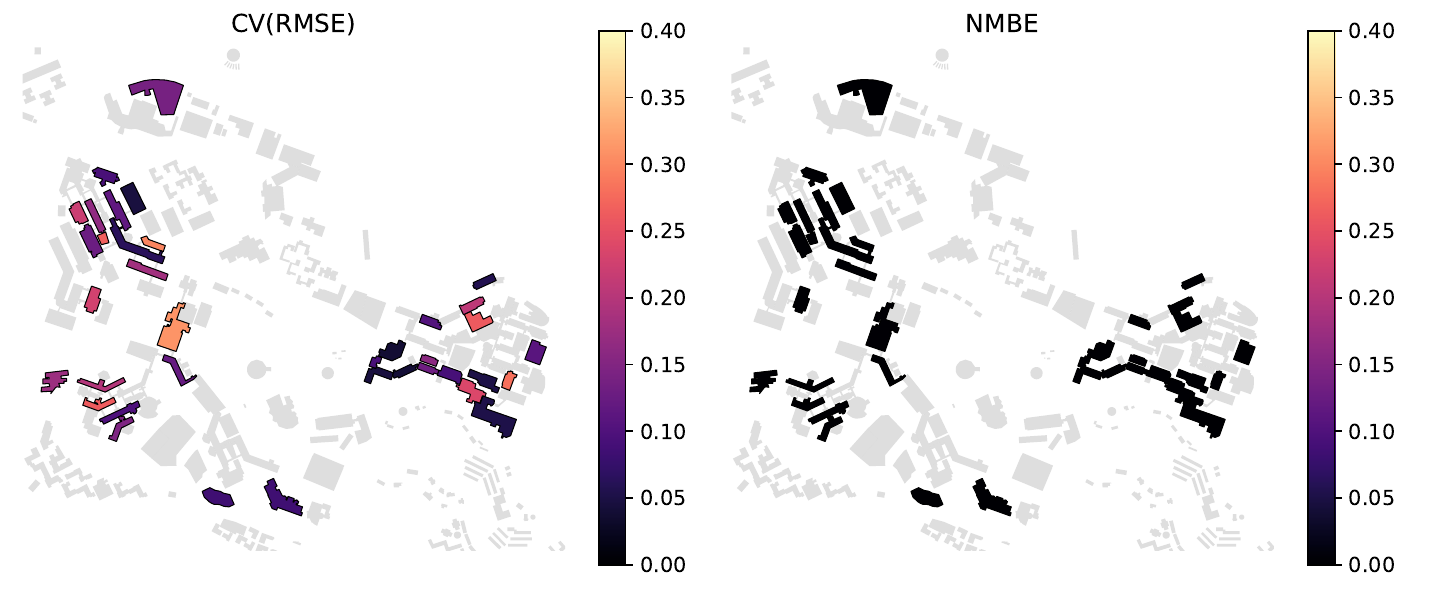}
  \caption{Mean hourly CV(RMSE) and NMBE in the electricity demand predicted by the regression model for each building in the case study with 10-fold cross-validation.}
  \label{fig:regression_map}
\end{figure}

The measured cooling demand data and WiFi counts were used in order to fit the regression model for the electricity demands for different end uses according to Eq.\ref{eq:electricity_tot}. The resulting electricity demands for three sample buildings are shown in Figure~\ref{fig:electricity_regression}. The regression model was tested by 10-fold cross-validation and the mean CV(RMSE) for each building was calculated. The resulting CV(RMSE) for each building is shown in Figure~\ref{fig:regression_map}. The regression model appears to be performing adequately for all buildings except for three, for which the CV(RMSE) is greater than 30\% at an hourly scale.

\subsection{Building occupant modeling and typical year input schedules} \label{sec:schedules}
In order to compare the future scenarios to a common baseline, an energy demand model for a ``typical'', pre-pandemic year was created. For this purpose, we created input schedules of occupancy and building system operation that represent the expected typical behavior of building occupants and systems as opposed to using measured data that might show unexpected, atypical deviations from standard behavior.

Profiles of building occupancy were derived from the number of devices connected to the WiFi network in each building at each hour of the three-year period considered. The WiFi connection profiles were Z-standardized and clustered by $k$-means clustering. Due to a change in the logging system, the dataset for 2020 is not comparable to the one for 2018--2019. Furthermore, building occupancy in the second half of 2020 was lower than in a normal year due to the COVID-19 pandemic. Hence, clustering was carried out separately for ``normal'' (pre-pandemic) years and 2020. 

\begin{figure}
  \includegraphics[width=\linewidth]{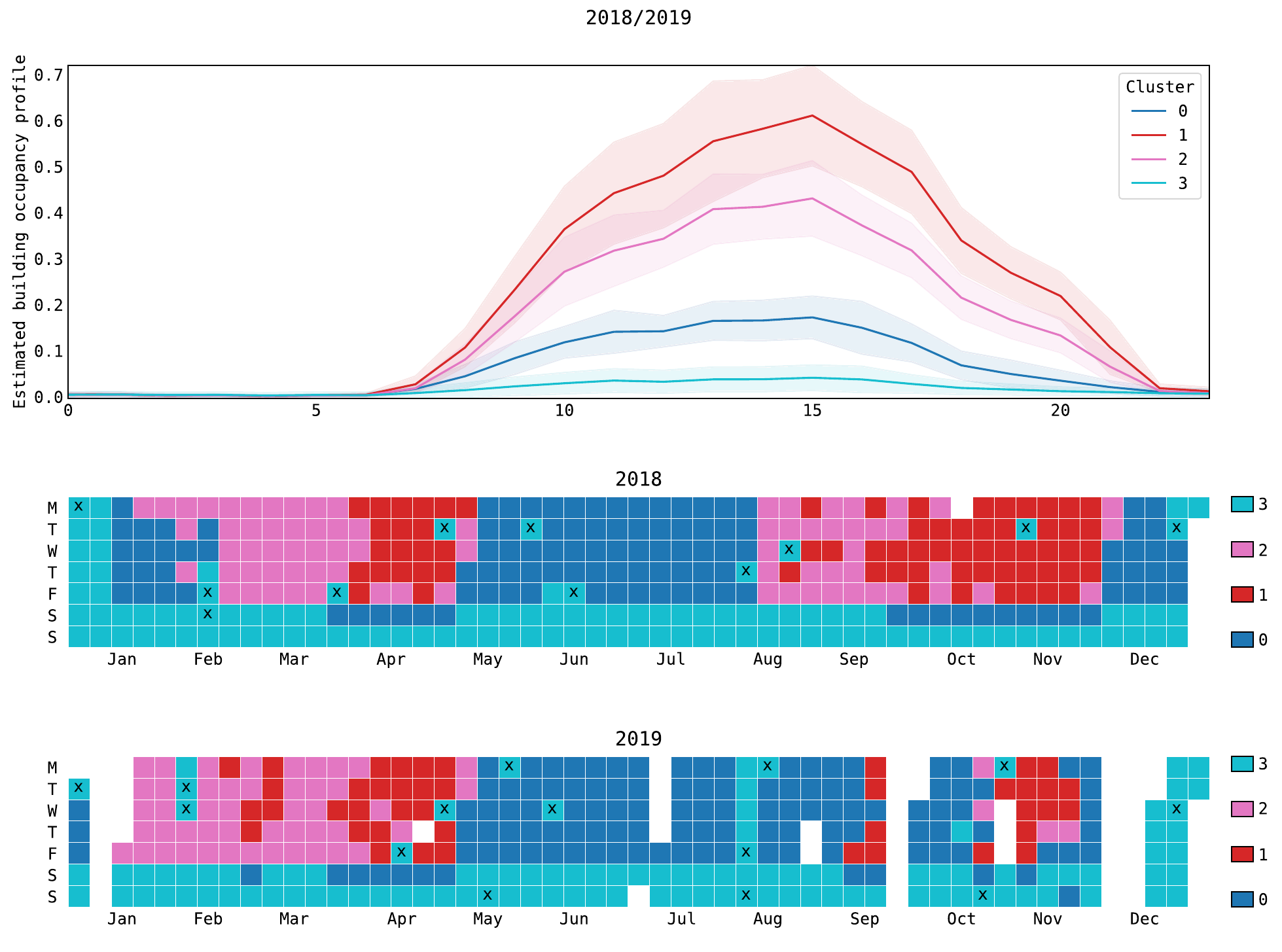}
  \includegraphics[width=\linewidth]{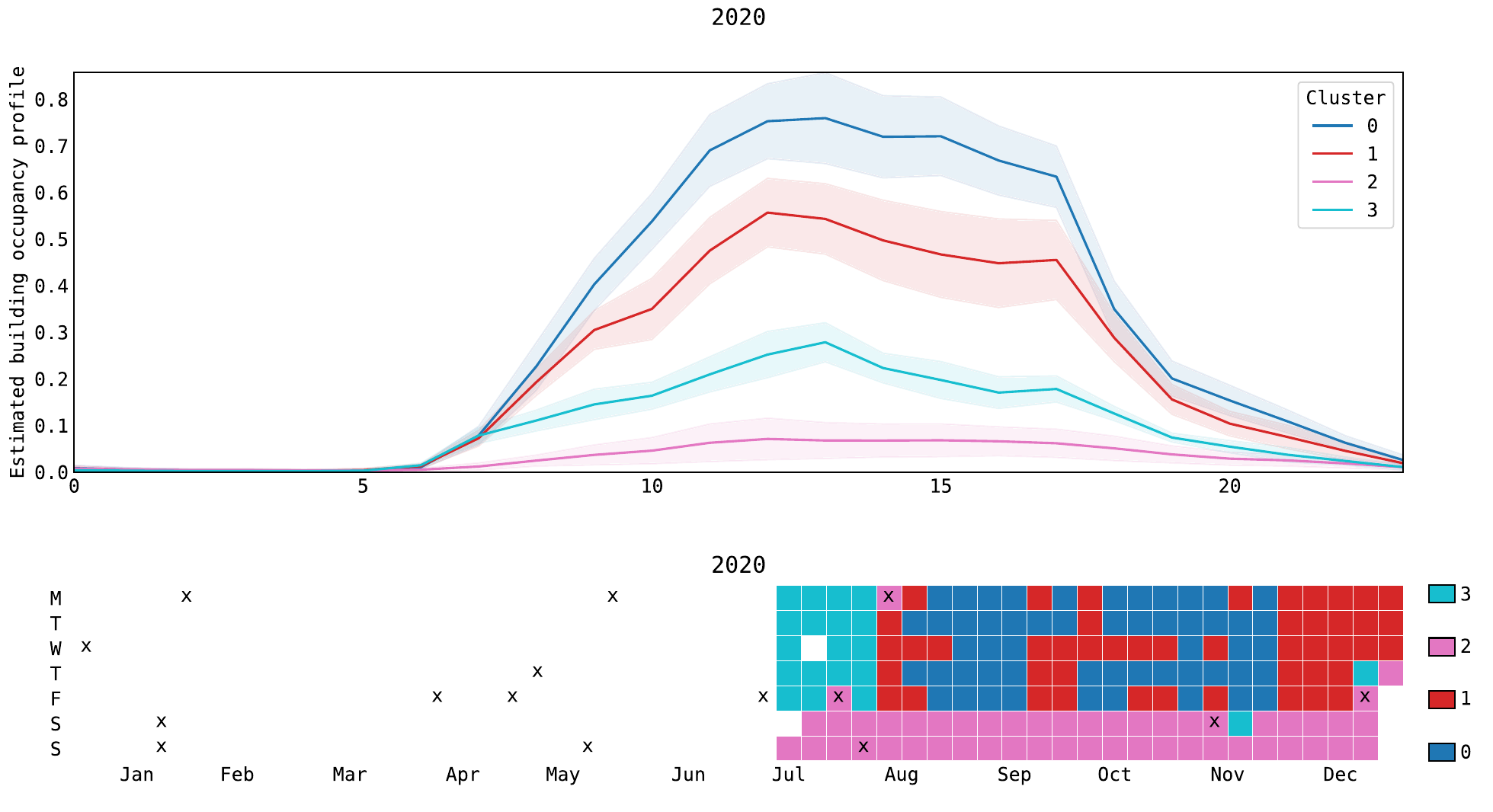}
  \caption{Clustering results for the WiFi data for a sample building for 2018 and 2019 (top) and 2020 (bottom). The average and standard deviation of the WiFi patterns for each of the years collected and the corresponding cluster number for each day are shown. Public holidays are shown as an X on the calendars.}
  \label{fig:schedules}
\end{figure}

Each day in the dataset was assigned an occupancy pattern cluster, and subsequently a ``weekday'', ``Saturday'' and ``Sunday'' schedule was assigned to each week in the year for each building in the area. Each day of the year was then assigned an occupancy pattern derived from the mean of the corresponding assigned cluster. An example of a building in the case study area is shown in Figure~\ref{fig:schedules}. The schedule consists of an hourly value between 0 and 1, representing the share of the maximum number of occupants in each building. As can be observed in Figure~\ref{fig:schedules}, the typical year occupancy may never reach the actual maximum occupancy of a building, which is common in a real office and classroom buildings. The clusters show the expected differences between weekdays and weekends, as well as a notable seasonal variability due to the academic calendar, with peak occupancy in February--April and August--November, during the peak of the academic semester, and lower occupancy during the summer and winter breaks. 

The most common weekday and weekend occupancy cluster for each building for each week in the year was first identified. The mean schedule for each cluster was then assigned as the daily schedule for the corresponding day of the year. The result of this process is yearly occupancy schedules that differentiate between weekdays and weekends at different times of the year for each building in the area. The schedules for each of the occupancy scenarios in Figure~\ref{fig:scenario_schematic} were then created by adjusting the peak number of occupants in each scenario. This means that if a building in the case study area was assigned a maximum occupancy of 70\% (as seen in Fig.~\ref{fig:schedules}), a 50\% share of people working from home would lead to a maximum occupancy of 35\% for the scenario in question.

The electricity loads for lighting and occupant-related plug loads are estimated based on Eqs.~\ref{eq:electricity_occ} and \ref{eq:electricity_l}, respectively. The sensible and latent gains from occupants are derived from the CEA archetypal values, in this case, 10 Watts per person and 80 grams of moisture per hour per person, respectively. The ventilation systems are assumed to be occupant-controlled, with a minimum ventilation rate of 10 liters per person per second, again based on the CEA archetypes.

Finally, binary cooling system operation schedules for each building for the baseline case were created based on the measured space cooling demand. For simplicity, the space cooling systems in the baseline case were assumed to be on at all hours of the day in which the cooling load exceeded the base load for more than 50\% of the hours of the year. The cooling system setpoint temperature $T_{cs}$ at each hour of the day $h$ was then defined as follows:
\begin{equation}
    T_{cs} (h) = 
        \begin{cases}
            T_{setpoint},  & \text{if } \frac{\sum_{t'}1}{\sum_{t}1} \geq 0.5~\forall~t' \in \tau(Q_{cs}(t) > Q_{cs,base})~\forall~t \in \tau(h) \\
            T_{setback},   & \text{otherwise}
\end{cases}
\end{equation}
where $\tau(h)$ is the set of time steps where the hour of the day was $h$, $t'$ is every time step where the measured cooling demand $Q_{cs}$ exceeded the cooling base load $Q_{cs, base}$, and $\tau$ is the set of all time steps $t$ in which the hour of the day is $h$. So if, for example, the cooling load at 8 in the morning is higher than the base load for 50\% of the days of the year, the cooling systems are assumed to be on at this hour for every day of the year. 

\section{Results} \label{sec:results}
\begin{figure}[ht]
  \flushright
  (a)
  
  \includegraphics[width=\linewidth]{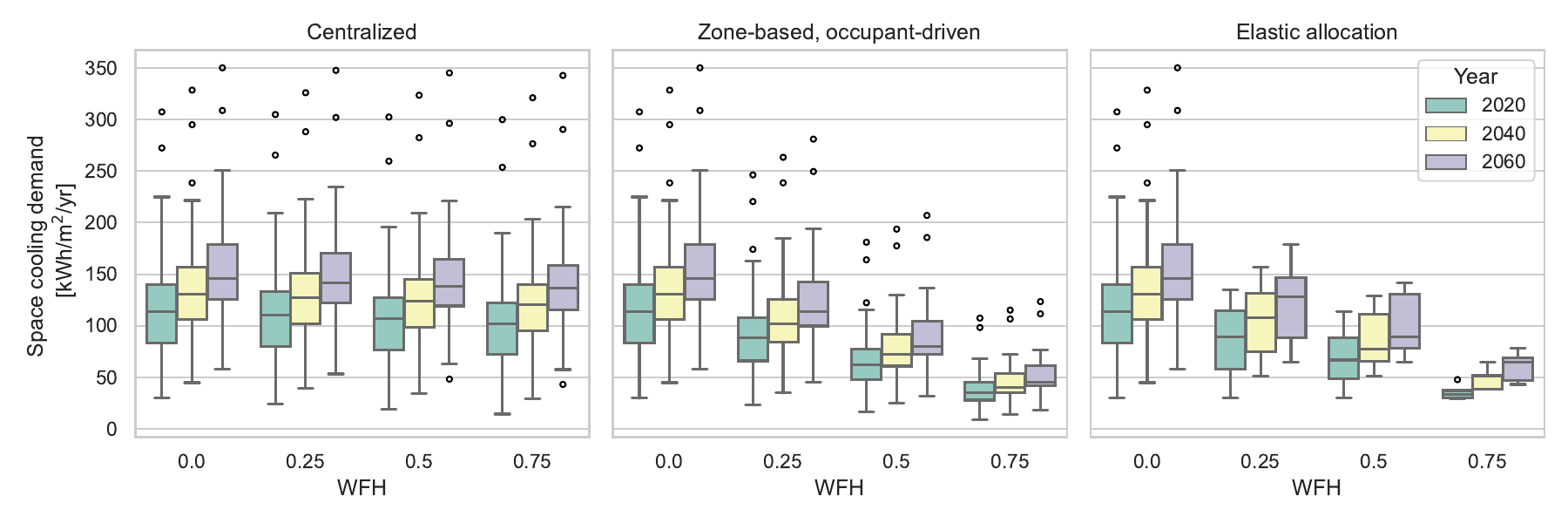}
  (b)
  
  \includegraphics[width=\linewidth]{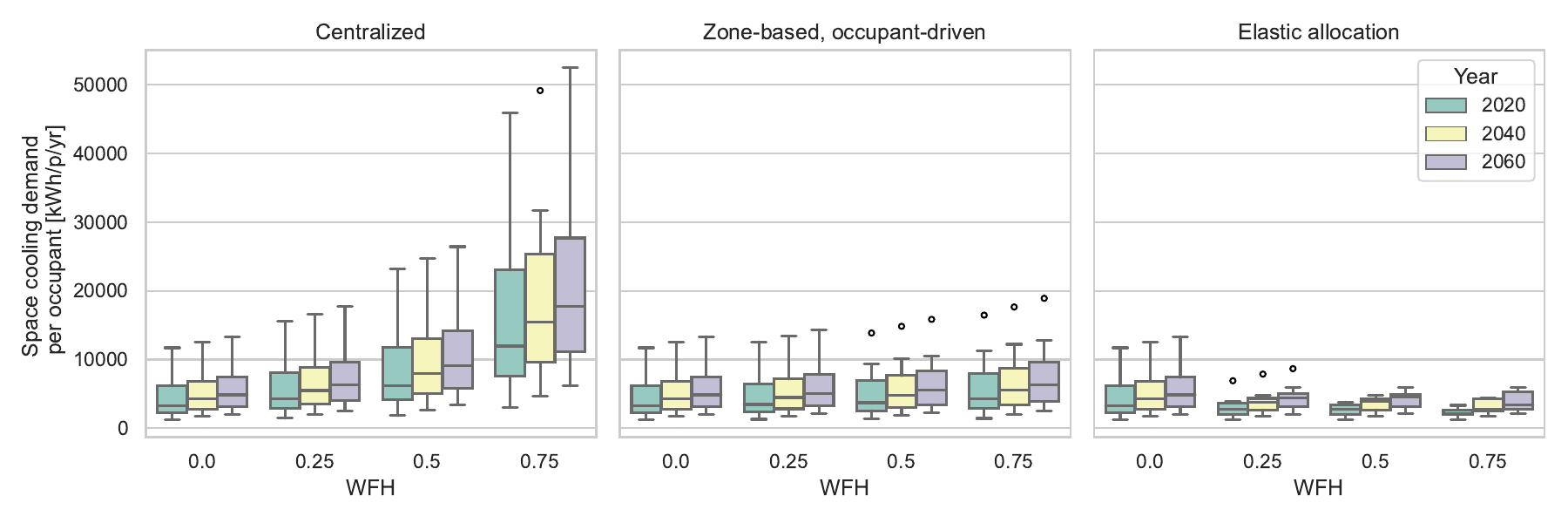}
  \caption{Yearly space cooling demand per m$^2$ (a) and per occupant (b) for each of the work-from-home (WFH) scenarios under three different building operation scenarios.}
  \label{fig:cooling_per_m2_per_capita}
\end{figure}

Figure~\ref{fig:cooling_per_m2_per_capita} shows the distribution of the space cooling demands for all buildings for each WFH and building operation scenario. The results for the first building operation scenario, where no changes are introduced to building operation, show that a decrease in building occupancy as a result of an increased reliance on remote working and studying post-COVID-19 would lead to a minimal decrease in the building cooling demand for the buildings in the case study area. For a reduction in building occupancy of 25--75\%, a decrease in space cooling demand of only 5--15\% at campus scale is observed. Furthermore, the effects of climate change are significant enough to make up for any savings observed due to reduced building occupancy (15\% higher total space cooling demand by 2040, 29\% by 2060). Maintaining the status quo in building operations would therefore lead to large amounts of energy waste if an increased share of home-based work and schooling becomes the norm moving forward.

A zone-based, occupant-driven building operation, where only the floor area that is actually occupied is assumed to be conditioned, unsurprisingly leads to a much more significant reduction in the space cooling demand, proportional to the decrease in building occupancy. Therefore, by introducing such dynamic building system controls, the space cooling demand could be significantly reduced. This effect is even more dramatic when considering the space cooling demand per building occupant, as seen in Figure~\ref{fig:cooling_per_m2_per_capita}. When no change to building operation is introduced, the demands per occupant rise sharply due to decreasing occupant density, whereas adjusting building operation accordingly allows the space cooling demand per occupant to be maintained within a fairly constant range.

A decrease in the number of people physically attending their offices and classrooms could potentially create an opportunity to re-develop the area's building stock and prioritize the operation of more energy-efficient buildings. In the final scenario, occupants 
are assumed to be relocated in order to ensure full building occupancy, even in cases of increased shares of remote working and studying. As seen in Figure~\ref{fig:cooling_per_m2_per_capita}, the outliers observed in the case with no remote working are eliminated, and instead more efficient buildings are prioritized, thus further reducing the space cooling demands in the case study area.

\begin{figure}
  \includegraphics[width=\linewidth]{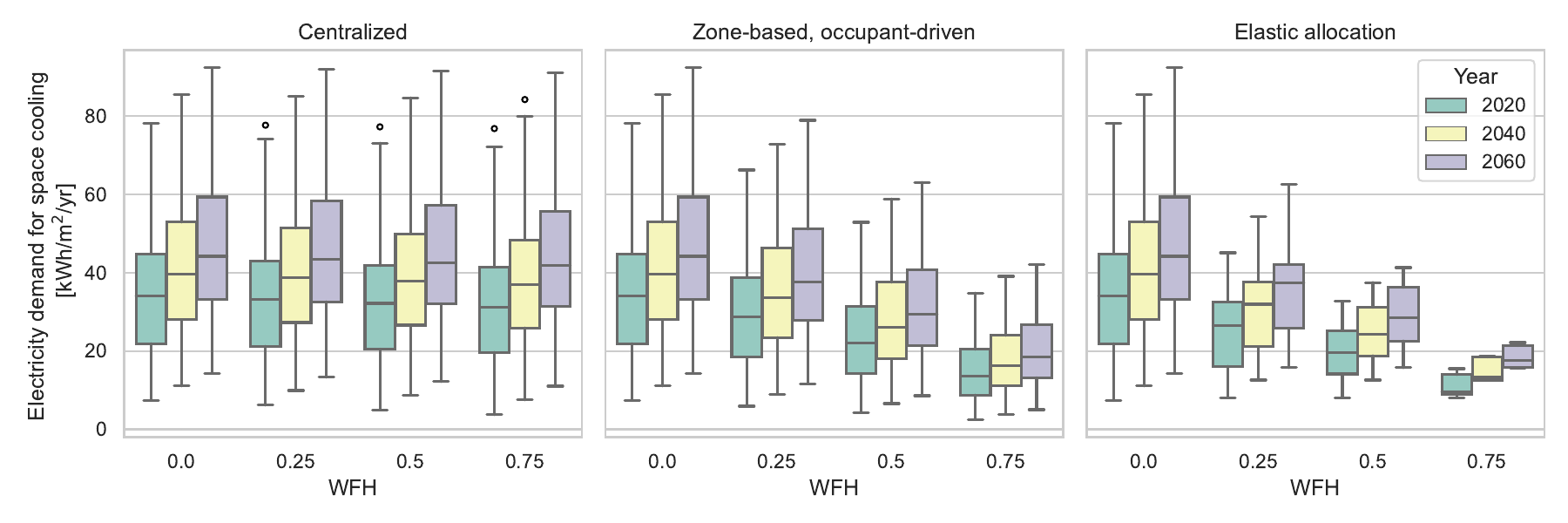}
  \caption{Yearly electricity demand for space cooling per m$^2$ for each of the work-from-home (WFH) scenarios under three different building operation scenarios.}
  \label{fig:electricity_per_m2}
\end{figure}

The yearly electricity demands per building for space cooling for each scenario are shown in Figure~\ref{fig:electricity_per_m2}. The results show that the electricity savings for the occupant-centric case are indeed slightly less pronounced than the space cooling savings due to the part-load operation of the building cooling systems. Nevertheless, the decreased system performance is not significant enough to revert the general trends observed in Figure~\ref{fig:cooling_per_m2_per_capita}. The electricity savings are most significant for the case of elastic space allocation, as all buildings are operated at full capacity while still considerably reducing the conditioned floor area of the campus.

\begin{figure}
  \includegraphics[width=\linewidth]{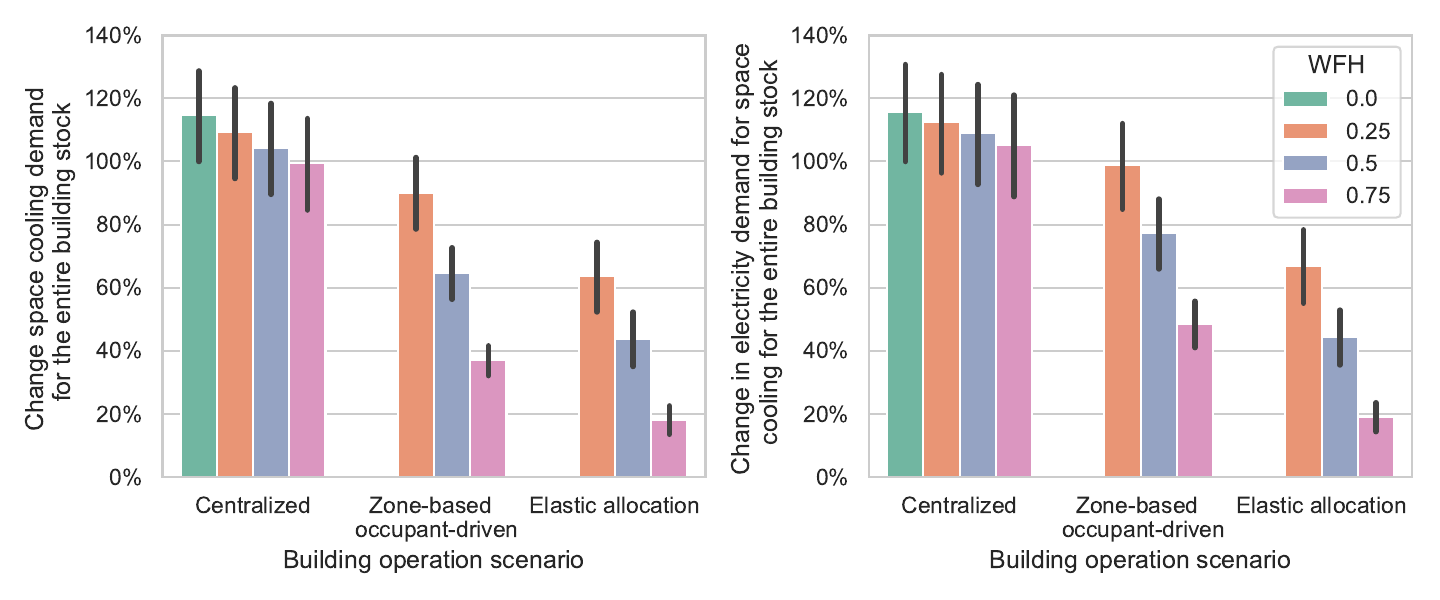}
  \caption{Change in yearly space cooling demand (left) and electricity demand for space cooling (right) for each of the work-from-home (WFH) scenarios under three different building operation scenarios. The error bars show the range of values obtained due to the effects of climate change from 2020 to 2060.}
  \label{fig:cooling_per_year_campus}
\end{figure}

The effect of part load cooling system operation on the demands for the entire building stock is shown in Figure~\ref{fig:cooling_per_year_campus}. Here, the simulation results for the case with 100\% occupancy and 2020 weather are used as a baseline for comparison of all scenarios for both space cooling and electricity demand. These results show that for the case with no change in building operation, the electricity savings due to decreased occupancy are even smaller than the savings in space cooling due to part load operation. The effects of part-load operation are marginal for the case with elastic space allocation, as buildings are operated at full capacity. For the zone-based occupant-driven controls, on the other hand, the space cooling savings compared to the centralized building operation strategy of 17--63\% observed (for WFH shares of 25--75\%) translate to 12--54\% savings in electricity demand. As previously mentioned, however, the effects of part load operation are not significant enough to revert the conclusion that such a building operation could, in principle, lead to lower energy demand for cooling. For the elastic space allocation case, since buildings are mostly operated at full occupancy, there are no effects of part load operation, and hence the reduction in space cooling and corresponding electricity demand is within the same range (40-84\% for WFH shares of 25--75\%). 

\begin{figure}[!h]
  \includegraphics[width=\linewidth]{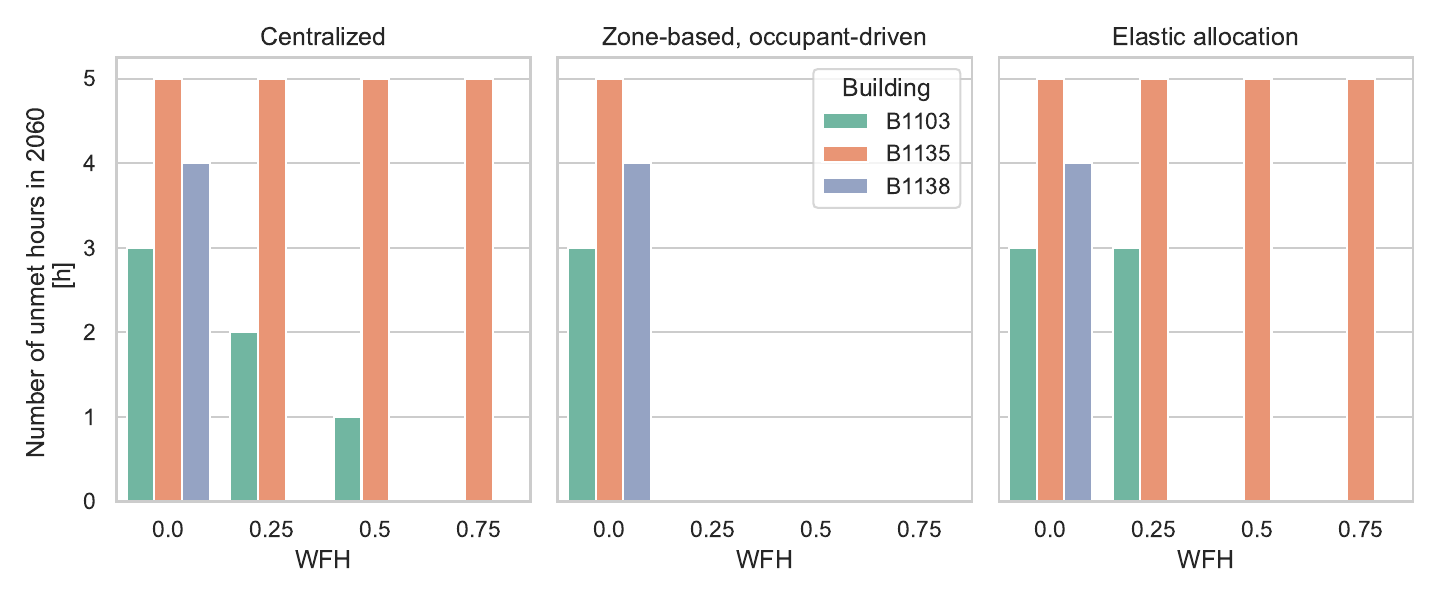}
  \caption{Number of hours in 2060 where the space cooling demand for three of the buildings in the case study cannot be met by the existing cooling systems for each scenario.}
  \label{fig:unmet_hours}
\end{figure}

Finally, as the space cooling demand increases due to the effects of climate change, the existing installed cooling capacity might eventually not be enough to meet the demand for space cooling in the case study buildings. For three buildings in the case study area, the installed cooling capacity was insufficient by 2060 in the baseline case, as shown in Figure~\ref{fig:unmet_hours}. For B1135, even at very low occupancy, the systems were unable to cope with future demands for a few hours per year. 
This problem is still observed in the WFH scenarios with elastic space allocation, as the buildings that are allowed to operate do so at full occupancy. In the case of zone-based, occupant-driven control, on the other hand, the demand is met for all buildings and for all hours of the year in each of the WFH scenarios. Thus, there is a tradeoff between decreased district-scale space cooling demand and optimized use of space in the elastic space allocation scenario, and increased system resilience in the zone-based, occupant-driven scenario.

The number of unmet hours is small in all cases, and in real-life situations, such a building operation might be acceptable for facility managers. Nevertheless, the trends shown demonstrate that climate change might indeed have an impact on the operation of existing building systems, even if relatively small. Under such circumstances, an occupant-driven building operation might help cope with the effects of climate change.

\section{Implications on the design of future district energy systems and campus operation} \label{sec:discussion}
The simulation results show the potential for occupant-driven building system controls and elastic building operations as an energy-saving strategy as flexible work arrangements become the norm in the coming years. Such systems could furthermore contribute to increasing the robustness of buildings to climate change, as they could allow buildings to cope with future demands without the need to increase the installed cooling capacity in buildings while providing the flexibility to meet the demands for cooling even if buildings were to be fully occupied on occasion. The negative effect of the part-load operation was not large enough to compensate for the additional cooling loads created by traditional centralized building system operation. Given that implementing these measures at the campus scale would require a significant shift in the way buildings and districts are operated, over the next sections, the types of technologies required and the implications of such planning decisions at the district scale are discussed critically.

\subsection{Technologies to support occupant-driven and elastic building operation}
Occupant-centric controls can be categorized into occupancy-based controls, which focus on the presence/absence of the occupant, and occupant behavior-based controls, which focus on occupant preferences that are inferred from occupants' interactions with building systems \cite{Park_etal_2019}. The modeling framework presented in this paper did not account for occupant preferences and interactions with building systems. However, the simulation baseline already assumed demand-controlled ventilation systems, whereby the amount of air supplied was directly proportional to the number of occupants present (as discussed in Section~\ref{sec:schedules}). Nevertheless, since no additional occupant interactions with building systems were considered, the building system controls considered in this paper fall into the occupancy-based category.


In this paper, we have used WiFi connection data as the basis for a building occupancy model. Given the ubiquity of WiFi-connected personal devices, such an approach could also be used for an occupant-driven building operation. WiFi has been found to be particularly well-suited for HVAC control and operation due to its potentially high occupancy and spatial resolution, high accuracy, and low costs \cite{Azimi_OBrien_2022}. 
By using existing IT infrastructure, the costly installation of sensing technologies can be avoided, a particularly appealing feature due to the difficulty and costs of sensor installation at the campus scale. 
Due to the limited spatial resolution of the CEA building models, the data used in this study were aggregated at the building level. However, since access points are located on every floor of the case study buildings, a more detailed description of occupant location based on WiFi devices could be achieved. Real-time location detection of WiFi-enabled devices on a given floor may be achieved by triangulation \cite{Wang_and_Shao_2017}, though such an approach may require additional nodes to be installed.


Mobile applications could furthermore play a key role in occupant-driven and elastic building operation. On the one hand, a system to book workspaces in advance (similarly to e.g., \cite{Sood_etal_2020}) could play a key role in optimizing the use of workspaces at the campus scale as suggested in the elastic space allocation scenario. Furthermore, wearable applications could be used to crowd-source subjective occupant comfort feedback from building occupants \cite{Sood_etal_2019}. Such subjective comfort feedback would help develop the types of learning-based, personalized comfort models required to operate occupant-driven building controls that minimize energy demand while maintaining acceptable comfort levels for building occupants.

Even when accounting for improved personalized thermal comfort models, a key issue in occupant-driven system operations is the differences in thermal preferences among the occupants in shared spaces \cite{Kim_etal_2018}. One potential solution to this issue is the installation of personalized conditioning systems, which aim to condition individual occupants' immediate surroundings without affecting others in the same space. A review of these technologies found that thermal comfort can be maintained at ambient temperatures that are 4--5$^\circ$C higher than the temperatures recommended by current standards in cooling-dominated climates \cite{Vesely_Zeiler_2014}. Furthermore, Anand et al. \cite{Anand_etal_2022} suggest personalized conditioning systems could help mitigate the risk of viral loads in confined spaces by supplying adequate ventilation rates at an increased zone setpoint temperature while maintaining thermal comfort and perceived air quality through elevated air movement. Thus, in addition to saving energy and improving robustness to climate change, an occupancy-driven HVAC system control coupled with personalized conditioning systems may help build resilience to the effects of future pandemic events.

\subsection{Further district planning considerations}
The lowest overall space cooling demands observed in the present case study correspond to the case of elastic space allocation, where inefficient buildings were completely closed off rather than optimizing the operation of partially-occupied buildings. By ensuring that buildings are only operated at full occupancy (and thus maximizing the performance of the cooling systems), the space cooling demands per occupant in the elastic space allocation case were minimized. While such a case may be somewhat idealized (as, for example, research spaces are not interchangeable), flexible workspaces may need to become the norm in order to minimize energy demand if working and learning from home indeed continue to be the norm at least part-time moving forward. 

While not part of the scope of the present study, the social implications of such a shift towards a fully flexible workspace allocation are worth keeping in mind. Marzban et al. \cite{Marzban_etal_2022} argue that introverted and orthodox individuals not only have a bigger struggle to adapt to this new way of working, but they might also undergo higher levels of psychological discomfort. They also cite previous studies that found several workers in offices supporting activity-based workspaces do not switch their workstations frequently; that is, occupants might have a tendency to reserve desks as their own even in flexible work environments. Hence, it might be worth considering whether the flexible working paradigm might be pushing out certain types of users from the office rather than ensuring all employees have access to a variety of work environments.

The positive implications on energy demand and environmental impacts might make this shift a necessity, regardless of its social implications. 
Achieving global decarbonization targets will require proactive policies such as supporting working from home and teleconferencing to reduce flying and commuting, especially if combined with the rationalization and reduction of office, administration, and other workspaces \cite{Kikstra_etal_2021}. It is therefore worth questioning whether university campuses might be ``oversized'' for the future needs of the researchers and students who occupy them. An increased reliance on working from home and flexible working and studying arrangements may support the transition to a more rational use of space in university campuses.

In addition to the energy-saving potential in such arrangements, there is an obvious benefit to organizations in saving on real estate and building operation costs \cite{Sood_etal_2020}. While clearly not the priority of an educational and research organization as considered in the present case study, a case could be made for a revenue-generating model in rationalizing office space on campus and renting out unused spaces to generate income as suggested by Yang et al. \cite{Yang_etal_2019}, potentially using the unoccupied square footage to host young companies and startups. In order to ensure that emissions are not merely passed on to the tenants, these buildings (which in the present study were assumed to be the least efficient in the dataset) would need to be appropriately retrofitted.

\section{Conclusions} \label{sec:conclusions}
This paper investigated the need for elastic building operation strategies to be implemented to support the shift towards increasingly flexible work arrangements and remote working and studying on a university campus in Singapore. The effects of flexible work arrangements on the space cooling demand at the campus scale were assessed through a calibrated district-scale energy demand simulation and different building operation scenarios. 

The results showed that building occupancies of 25\% to 75\% due to flexible work arrangements led to minimal space cooling savings (5--15\%) if buildings were operated in a conventional, centralized way. Introducing occupancy-driven, zone-based building controls could lead to significant savings in space cooling demand (17--63\%) with respect to the centralized baseline at building occupancy shares of 25--75\%. The greatest decrease in space cooling demand was observed for the case with an elastic space allocation approach, where occupants were assumed to be relocated in order to minimize the operation of the most inefficient buildings in the case study area (40--84\% for the same occupancy shares).

The effects of part-load operation for the decentralized, occupant-driven case were not found to be significant enough to affect the overall conclusions supporting flexible building system operation strategies. Furthermore, the long-term effects of climate change were also considered, and were found to potentially lead to some space cooling demands not being met for the centralized and elastic allocation cases as soon as 2040. The zone-based, occupant-driven approach, on the other hand, was able to meet the loads for all cases, as systems did not operate at full capacity in the work-from-home scenarios.

The results stress the importance of introducing flexible operation strategies into buildings to achieve energy savings if flexible work arrangements continue to be pursued moving forward. A number of building systems to support this transition were introduced, focused mostly on occupant detection and zone-based building system operation. As the lowest demands in the case study area were achieved in the case of elastic space allocation, however, it is worth questioning whether there will also be a need to reduce and rationalize workspaces to achieve future decarbonization targets in university campuses.

\subsection{Limitations and future work}
The methodology presented in this paper was envisioned from a district scale perspective. Therefore, the methodology relies on relatively simple inputs and data sources, where only building energy metering and WiFi data are available. However, this translates to a number of simplifying assumptions. In the absence of detailed building floor plans, material specifications, and system operation data, model calibration was used in order to produce more reliable estimates. In order to further improve model accuracy, measurement campaigns and site visits for each of the buildings in the case study could be carried out, although this might still be impractical for larger case studies.

Another simplifying assumption in the definition of the elastic allocation scenario was that workspaces within the campus were essentially interchangeable. That is, occupants in one building use type could be relocated to any other building that contained the same use type. This might not be true for real workspaces on campus, particularly for laboratories, which can differ greatly depending on each discipline's needs (i.e., a physics laboratory has very different requirements from a biology laboratory). This means that the level of workspace optimization assumed in this work might not be feasible in a real campus, leading to lower energy savings than predicted in this study. Nevertheless, the positive impacts observed point to the need to make workspaces flexible wherever possible in order to minimize the energy demands of such spaces if working and learning from home indeed continue to be the norm, at least part-time, moving forward. 

As acknowledged in Section~\ref{sec:discussion}, the shift in building operation presented in this paper will have significant implications on the technologies implemented as well as on planning considerations at the campus scale. In terms of technologies, we have sought to propose technologies that require minimal intervention at campus scale (i.e., pre-existing WiFi networks, application-based space allocation, etc.). However, the social implications on building occupants and their ability to carry out their activities in such arrangements would require further research.

From this perspective, another simplifying assumption in our work has been that the share of people working and studying on campus in each scenario and the allocation of spaces within the district to them was defined in a top-down manner. In order to plan and assess occupant-driven building and campus operation approaches, occupant-driven modeling approaches that can account for the needs of different occupant types are also required. Thus, we are currently working on an agent-based model of building occupants at the campus scale such that our building operation scenarios could be driven by occupant behavior and thermal comfort rather than defined being defined in a top-down manner as presented in this paper. 

\section*{CRediT authorship contribution statement}
\textbf{Mart\'in Mosteiro-Romero:} 
    Conceptualization, 
    Data curation, 
    Formal analysis, 
    Methodology, 
    Project administration, 
    Validation, 
    Visualization, 
    Writing -- original draft. 
\textbf{Clayton Miller:} 
    Conceptualization, 
    Methodology, 
    Project administration, 
    Resources, 
    Supervision, 
    Writing -- review \& editing.
\textbf{Adrian Chong:} 
    Data curation, 
    Methodology, 
    Writing -- review \& editing.
\textbf{Rudi Stouffs:} 
    Conceptualization, 
    Funding acquisition, 
    Project administration, 
    Resources, 
    Supervision, 
    Writing -- review \& editing.
    
\section*{Acknowledgments}
The research was conducted at the Singapore-ETH Centre, which was established collaboratively between ETH Z\"urich and the National Research Foundation Singapore. This research is supported by the National Research Foundation, Prime Minister's Office, Singapore, under its Campus for Research Excellence and Technological Enterprise (CREATE) program.

The authors would like to thank Sicheng Zhan for his assistance in collecting and preprocessing the building energy demand and WiFi data.

\section*{Data availability}
The research compendium for this article is available on a public GitHub repository:
\newline \url{https://github.com/buds-lab/elastic-buildings}.

\bibliography{main}

\end{document}